\documentclass[prb,superscriptaddress,twocolumn]{revtex4}

\usepackage{graphicx,color}
\usepackage{amsmath}
\usepackage{amssymb}
\usepackage[normalem]{ulem}
\usepackage[colorlinks=true,citecolor=blue]{hyperref}

\usepackage{bbold}
\usepackage{tikz}
\usetikzlibrary{patterns}

\newcommand{\bracket}[2]{\langle#1|#2\rangle}

\newcommand{\ri}{{\rm i}}

 \newcommand\beq            {\begin{equation}}
  \newcommand\eeq          {\end{equation}}
 \newcommand\bwt         {\begin{widetext}}
 \newcommand\ewt         {\end{widetext}}

\newcommand{\pt}{\tilde{p}} 
\def\e{{\rm e}}

\usepackage{times}
\usepackage{amssymb}
\usepackage{amsfonts}
\usepackage{amsmath}
\usepackage{graphicx}
\usepackage{bm}
\usepackage{enumerate}

\def\tfract#1/#2{{\textstyle{\raise0.8pt\hbox{$\scriptstyle#1$}\over%
\hbox{\lower0.8pt\hbox{$\scriptstyle#2$}}}}}

\def\radi2k{\tfract 1/{\sqrt {2k}} }
\def\der{\partial}
\def\downnormalfill{$\,\,\vrule depth4pt width0.4pt
\leaders\vrule depth 0pt height0.4pt\hfill\vrule depth4pt width0.4pt\,\,$}
\def\WT#1{\mathop{\vbox{\ialign{##\crcr\noalign{\kern3pt}
      \downnormalfill\crcr\noalign{\kern0.8pt\nointerlineskip}
      $\hfil\displaystyle{#1}\hfil$\crcr}}}\limits}
\def\be{\begin{equation}}
\def\ee{\end{equation}}
\def\bes{\begin{equation*}}
\def\ees{\end{equation*}}
\def\bea{\begin{eqnarray}}
\def\eea{\end{eqnarray}}
\def\beas{\begin{eqnarray*}}
\def\eeas{\end{eqnarray*}}
\def\ba{\begin{array}{rcl}}
\def\ea{\end{array}}
\def\der{\partial}
\numberwithin{equation}{section}
\def\go{\leavevmode \raise.3ex\hbox{$\scriptscriptstyle \langle\!\langle\!  $}%
~\ignorespaces}

\def\gf{\relax \ifhmode \unskip~\else \leavevmode \fi \raise.3ex\hbox{$\! \scriptscriptstyle\rangle\!\rangle\, $}}
\begin{document}

\author{Michele Burrello}
\affiliation{Center for Quantum Devices and Niels Bohr International Academy, NBI, University of Copenhagen, Lyngbyvej 2,
2100 Copenhagen, Denmark.}

\author{Enore Guadagnini}
\affiliation{Dipartimento di Fisica E. Fermi, Universit\`a  di Pisa, Largo B. Pontecorvo 3, 56127 Pisa, Italy.}
\affiliation{INFN, Sezione di Pisa, Largo B. Pontecorvo 3, 56127 Pisa, Italy.}

\author{Luca Lepori}
\affiliation{Istituto Italiano di Tecnologia, Graphene Labs, Via Morego 30, I-16163 Genova, Italy.}
\affiliation{Dipartimento di Scienze Fisiche e Chimiche, Universit\`a dell'Aquila, via Vetoio,
I-67010 Coppito-L'Aquila, Italy.}
\affiliation{INFN, Laboratori Nazionali del Gran Sasso, Via G. Acitelli,
22, I-67100 Assergi (AQ), Italy.}

\author{Mihail Mintchev}
\affiliation{Dipartimento di Fisica E. Fermi, Universit\`a  di Pisa, Largo B. Pontecorvo 3, 56127 Pisa, Italy.}
\affiliation{INFN, Sezione di Pisa, Largo B. Pontecorvo 3, 56127 Pisa, Italy.}

\title{Field theory approach to the quantum transport in Weyl semimetals}

\begin{abstract}

We analyze the structure of the surface states and Fermi arcs of Weyl semimetals as a function of the boundary conditions parameterizing the Hamiltonian self-adjoint extensions of a minimal model with two Weyl points. These boundary conditions determine both the pseudospin polarization of the system on the surface and the shape of the associated Fermi arcs. We analytically derive the expectation values of the density profile of the surface current, we evaluate the anomalous Hall conductivity as a function of temperature and chemical potential and we discuss the surface current correlation functions and their contribution to the thermal noise. Based on a lattice variant of the model, we numerically study the surface states at zero temperature and we show that their polarization and, consequently, their transport properties, can be varied by suitable Zeeman terms localized on the surface. We also provide an estimate of the bulk conductance of the system based on the Landauer-B\"uttiker approach. Finally, we analyze the surface anomalous thermal Hall conductivity and we show that the boundary properties lead to a correction of the expected universal thermal Hall conductivity, thus violating the Wiedemann-Franz law.
  
\end{abstract}

\maketitle

\section{Introduction}

\textcolor{black}{Weyl semimetals are the focus of intense theoretical (see, for example, the reviews \onlinecite{hosur13,hasan17,yan17,burkov18,vishwanath18}) and experimental \cite{lv2015,weng2015,xu2015,xu2015_5,xu2015_3,huang2015,zhang2016} studies. They} are the prototypical example of three-dimensional systems that, despite being gapless in the bulk, display topologically protected properties \cite{wan11,balents11}. These properties stem from the chiral behavior of the Weyl band-touching points characterizing these materials, and they include peculiar transport phenomena such the chiral magnetic effect \cite{burkov2012} and the anomalous Hall response \cite{balents11,yang2011}, which are different manifestations of the quantum chiral anomaly that these systems display\cite{nielsen1983,burkov2012,huang2015,zhang2016,goswami2013,parame14,huang17,lepori2018}.

One of the most striking effects of the chirality of the Weyl points is the presence of gapless chiral states that are localized on the surfaces of Weyl semimetals and are protected against disorder. These chiral surface states appear along surfaces orthogonal to the separation of the Weyl points in momentum space, and they are responsible for an anomalous Hall conductivity proportional to the distance of the projection of the Weyl points on the surface Brillouin zone\cite{balents11,yang2011}.

Several experiments exploited ARPES techniques to detect these surface states \cite{xu2015_3} (see also \onlinecite{hasan17,yan17} and references therein) and the corresponding Fermi arcs \cite{wan11,balents11}; the shape and spin texture of the Fermi arcs, in particular, are non-universal features that strongly depend on the surface properties of the investigated materials. Despite being non-universal, however, the surface features play a fundamental role in defining some of the transport properties of Weyl semimetals: Ref. \onlinecite{cheianov18}, for instance, shows that the boundary characteristics are crucial in evaluating the surface contribution to the current induced by the chiral magnetic effect in an alternating magnetic field \cite{beenakker16,Pesin16,cheianov17}.

In this work, we present an analytical description of the surface states, currents and anomalous Hall conductivity of a minimal model of a Weyl semimetal.

Our results are based on the study of the self-adjoint extensions of the bulk Hamiltonian to the surface \cite{witten15,hashimoto2017,seradjeh18,faraei18}, which determine the set of physical boundary conditions describing the interface between the Weyl semimetal and the vacuum.
We consider only the ballistic regime, thus neglecting any disorder or interparticle scattering. This implies that our analysis neglects the dissipation effects from the surface to the bulk states (see, for example, Refs. \onlinecite{gorbar16,resta18} for an analysis of these effects on the surface transport), and additional disorder effects on the Fermi arcs \cite{juricic17}.

Our results provide an analytical description of the surface transport beyond linear response theory. In particular, we derive the profiles of the current density and conductivity as a function of the distance from the surface and we analytically evaluate the anomalous Hall conductance as a function of chemical potential, temperature and boundary conditions of the system. We calculate the thermal noise due to the surface states and we derive the anomalous Hall thermal conductivity of the system, finding that the Wiedemann-Franz law is fulfilled by the surface states only in the zero-temperature limit. 

We also verify numerically that it is possible to vary the boundary condition through suitable surface Zeeman interactions and, consequently, to change the value of the anomalous Hall conductivity of the system at non-vanishing chemical potentials. 

Overall, these results provide useful tools to interpret the surface transport properties of samples of Weyl semimetals in the ballistic regime at finite temperature and non-zero chemical potential, and to suitable estimate the effects of their boundary conditions.

This article is structured in the following way: in Sec. \ref{model} we introduce a low-energy description for a Weyl semimetal with two band-touching point; in particular, we emphasize the role of its boundary conditions and we construct a suitable quantum field theory for its study. In Sec. \ref{zeroT} its surface properties are studied and we present our main results about the anomalous Hall conductivity. Sec. \ref{bulkcond} provides an analytical estimate of the bulk conductance of the system. In Sec. \ref{num} we compare the zero-temperature results with the numerical study of its corresponding lattice model. Sec. \ref{secT} is devoted to additional properties of the system at finite temperature, focusing in particular on the surface thermal noise and anomalous thermal Hall conductivity. Finally we present our conclusions in Sec. \ref{concl}.

\section{The model and its boundary conditions} 
\label{model}

\subsection{The Hamiltonian and its spectral properties}

The Weyl points in a topological semimetal appear always in pairs with opposite chirality \cite{nielsen1981}. For this reason, a minimal model describing a realistic Weyl semimetal must include two band-touching points and break time-reversal invariance, as in the case of layered intermetallic materials with a trigonal crystal structure \cite{soh2019}. Our starting point is therefore a toy model of fermions with a suitable pseudospin-1/2 degree of freedom, that can represent orbital, sublattice or spin degrees of freedom.  The fermions move on a cubic lattice with a dynamics dictated by the Hamiltonian $H_{\rm lat} = \sum_{\bf p} c^\dag_{\bf p} H_{\rm lat}({\bf p}) c_{\bf p}$, where $c$ and $c^\dag$ are two-component spinors and:
\begin{multline} 
\label{hlat2}
H_{\rm lat}({\bf p}) = \tilde{v}\left(\cos p_0 - \cos p_x\right) \sigma_x +\\  v\left(2-\cos p_y - \cos p_z\right) \sigma_x +v \sin p_y \sigma_y+v \sin p_z \sigma_z\,.
\end{multline}
In this equation, the Pauli matrices $\sigma_i$ act on the pseudospin, and hereafter we adopt units such that the lattice spacing is unity.

This minimal model \cite{yang2011,okugawa2014,lopez2018} displays two Weyl points 
in $\left(\pm p_0,0,0\right)$ at zero energy, and we choose $\tilde{v}=v/\sin p_0$ in order to obtain an 
isotropic energy dispersion around both. For $\tilde{v}\cos p_0 \neq 0$, the Hamiltonian 
\eqref{hlat2} corresponds to a stack of 2-dimensional topological insulators laying on 
the $yz$ planes and coupled by the tunneling term along the $\hat{x}$ direction. 
In particular, in the two-dimensional $yz$ limit described by $\cos{p_x}=0$, we 
obtain a topological insulator with chiral gapless edge modes for $\tilde{v}\cos{p_0} < 0$. 
When introducing the coupling in the $\hat{x}$ direction, these gapless modes evolve 
into Fermi arcs localized on the surfaces $xy$ and $xz$.

The Hamiltonian \eqref{hlat2} is invariant under space-inversion symmetry,
\beq
H_{\rm lat}(-{\bf p}) = U \, H_{\rm lat}({\bf p}) \, U^{-1},  \quad \text{with } \; U = \sigma_x\, ,
\eeq
in such a way that the two Weyl points appear at the same energy. Concerning its boundaries, we neglect band-bending potentials at the surface; as a result, the Fermi arcs in this system do not display a spiraling dispersion, differently from the setups analyzed in Ref. \onlinecite{andreev15} (see also Ref. \onlinecite{lepori16} for the effect of trapping potentials).

To analytically study the behavior of the surface states, for small values of $p_0$, 
we approximate the low-energy behavior of \eqref{hlat2} with the Hamiltonian 
\begin{equation} \label{ham}
H({\bf p}) = \frac{{v}}{2p_0} \left(p_x^2-p_0^2\right)\sigma_x + vp_y \sigma_y + vp_z \sigma_z \, , 
\end{equation} 
leading to the differential operator 
\begin{equation} \label{ham1}
H(-i{\bf \nabla}) = -\frac{{v}}{2p_0} \left(\partial_x^2 + p_0^2\right)\sigma_x - v i \partial_y \sigma_y - v i \partial_z \sigma_z \, .  
\end{equation} 
Let us assume that the Weyl semimetal is located in the half-space ${\mathbb R}^3_+$ with $z>0$. Then (\ref{ham1}) 
defines a {\it symmetric} operator on a suitable set ${\cal D}({\mathbb R}^3_+)$ of smooth functions, 
which is dense in the set of square integrable functions on ${\mathbb R}^3_+$. 
The general theory (see e.g.  Ref. \onlinecite{RS}) of such operators implies in our case that $H(-i{\bf \nabla})$ has 
{\it self-adjoint} extensions, which involve {\it one} real parameter. 
The nature of this parameter can be deduced from the condition
\begin{equation} \label{selfadj}
\bracket{\psi H(-i{\bf \nabla})}{\varphi} - \bracket{\psi}{H(-i{\bf \nabla})\varphi} = 0 \,,
\end{equation}
imposed for each pair of wave functions $\psi$ and $\varphi$ in the domain ${\cal D}({\mathbb R}^3_+)$. 
The previous relation is equivalent to the surface condition
\be
\varphi^\dagger (\bm r)  \sigma_z  \psi(\bm r ) \,  \bigr |_{z =0}  = 0 \, ,
\label{boundary}
\ee
where ${}^\dagger$ stands for Hermitian conjugation. Analogously to the 2D case of graphene \cite{mccann2004,akhmerov2008}, Eq. (\ref{boundary}) 
corresponds to the physical requirement of vanishing of the probability current flowing across 
the surface $z=0$. It is satisfied  by the (maximal) set of wave functions with an arbitrary and 
translationally invariant polarization of the pseudospin parallel to the surface in all its points, namely 
\be
\left ( \sigma_x \cos \gamma + \sigma_y \sin \gamma \right )  \psi (\bm r )  \,  \bigr |_{z=0}  =  \psi (\bm r )  \,  \bigr |_{z =0} \, . 
\label{bc}
\ee
The angle $0\le \gamma <2\pi$ parametrizes all the self-adjoint extensions 
$H_\gamma $ of the Hamiltonian \eqref{ham} and
specifies the pseudospin polarization in the plane $xy$ of all the wavefunctions on the surface $z=0$. 
In the physical context the parameter $\gamma$ is expected to depend in general on the electric/magnetic properties of 
the Weyl material and the termination of its lattice that defines the surface, as experimentally 
verified in Refs. \onlinecite{souma2016,morali2019}.

Once $\gamma \in [0,2\pi)$ is fixed, the spectral properties of $H_\gamma $ are uniquely determined 
and concisely described in Appendix A. Summarizing, the eigenfunctions 
\be 
\{\zeta_{\rm s}^\pm ({\bf r}, p),\, \zeta_{\rm b}^\pm ({\bf r, p})\} 
\label{a}
\ee 
are of two different types, called in what follows {\it surface} and {\it bulk} states. The surface eigenstates 
$\zeta_{\rm s}^\pm ({\bf r}, p)$, given by (\ref{A1}), 
depend only on the two momenta $p =(p_x,p_y)$ parallel to the $z$-plane and 
decay exponentially along the $z$-axis. The bulk eigenstates $\zeta_{\rm b}^\pm ({\bf r, p})$, given by 
(\ref{A2},\ref{A3}), depend instead on all the three momenta ${\bf p}=(p_x,p_y,p_z)$ and oscillate along the $z$-axis. 
The associated eigenvalues 
\be 
\{\varepsilon_{\rm s}(p),\, \pm \varepsilon_{\rm b}({\bf p})\} 
\label{b}
\ee 
are expressed in terms of the combinations 
\be 
g(p_x) = \frac{p_x^2-p_0^2}{2 p_0} \, ,  \quad \tilde{p}_z(p)= p_y \cos \gamma - g(p_x) \sin \gamma \, . 
\label{g}
\ee
In particular, the surface states $\zeta_{\rm s}^\pm ({\bf r}, p)$ correspond to the domains in momentum space with positive and negative energies $\varepsilon_{\rm s}(p)$ respectively. These surface eigenvalues of the Hamiltonian are defined by:  
\be 
\varepsilon_{\rm s}(p) = v \left[g(p_x)\cos\gamma + p_y \sin\gamma \right]\, , \quad \tilde{p}_z(p)>0\, .
\label{s2}
\ee
The bulk energies result instead:
\be 
\varepsilon_{\rm b}({\bf p}) = v \sqrt{g(p_x)^2+p_y^2+p_z^2}\, , \quad  p_z{\ge}0\, .
\label{b2}
\ee
 The eigenvectors (\ref{a}) form 
a {\it complete orthogonal} basis, 
which together with (\ref{s2},\ref{b2}), defines uniquely\cite{RS} the self-adjoint extension $H_\gamma $.

\begin{figure}[t!]
\includegraphics[width=\columnwidth]{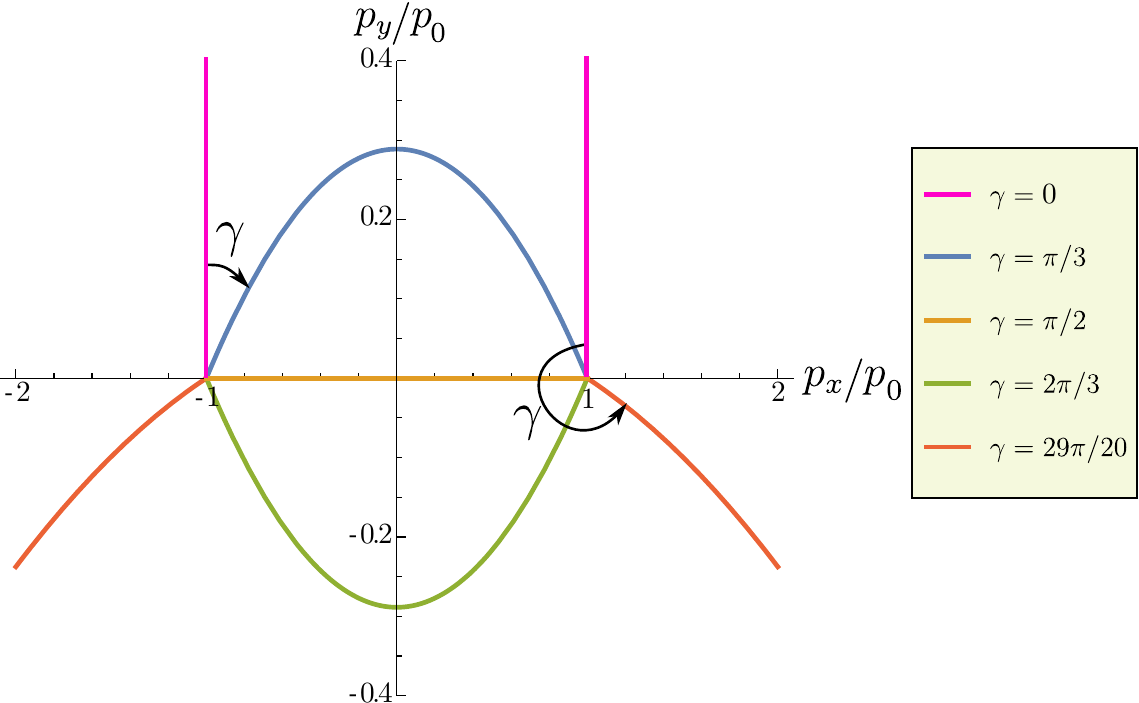}
\caption{The Fermi arc in the $xy$ surface is depicted for several values of $\gamma$ and $\mu=0$. 
The momenta are in units of $p_0$.}\label{fig:arcs}
\end{figure}  

The $\gamma$-dependence of the energy spectrum is a first indication that the physics 
of our system depends on the boundary conditions. It is instructive to consider in this respect the zero-energy eigenstates. There are four bulk states 
with this property, corresponding to the spin-degenerate states at the momenta $(p_x = \pm p_0, \,p_y =0 , \, p_z =0 )$. In addition, there is the family of surface states with vanishing energy 
\begin{equation}
p_y =  -g(p_x)\cot \gamma \, , \qquad \tilde{p}_z(p)>0 \, .  
\label{fa1}
\end{equation}
associated with the dispersion relation \eqref{s2}. 
The set of all these zero-energy eigenstates form 
the open Fermi arc on the surface Brillouin zone. In fact, equation \eqref{s2} defines the shape of the arc 
which depends explicitly on the angle $\gamma$, which gives its orientation at the limiting points $(p_x = \pm p_0, \,p_y =0 , \, p_z =0 )$ (measured from the upper vertical direction clockwise and counterclockwise respectively, see Fig. \ref{fig:arcs}). 
In particular, for $p_x= p_0$, the Fermi arc is always oriented orthogonally to the pseudospin polarization given by the boundary conditions \eqref{bc}. 
 In the range $0<\gamma<\pi$, the length $L_{\rm FA}(\gamma )$ of the Fermi arcs following from (\ref{fa1}) is 
\begin{equation}
L_{\rm FA}(\gamma ) = p_0 \left \{\frac{1}{\sin \gamma } + (\tan \gamma )\log \left [\frac{1+\cos \gamma}{\sin \gamma }\right ]\right \} \, . 
\label{fa} 
\end{equation}
We find Eq. \eqref{fa} instructive because
it relates the angle $\gamma$, which has an abstract mathematical
origin, with the length of the Fermi arc $L_{\rm FA}$ and with $p_0$, which 
are experimentally measurable quantities in ARPES measurements \cite{xu2015,xu2015_3,souma2016,morali2019}. 

The value $\gamma=\pi/2$ describes a straight Fermi arc connecting the projections of the 
Weyl points. In fact, at this point the function (\ref{fa}) reaches its minimum $L_{\rm FA}(\pi/2 ) = 2p_0$. 
For $\gamma =0$ and $\gamma=\pi$ the orientation of the Fermi arc 
becomes orthogonal to the line connecting the Weyl points: in this case and, 
more in general, for $\pi \le \gamma \leq 2\pi$, our choice of the function $g$ 
returns indeed two distinct and unbounded Fermi arcs. In tight-binding models of real materials, the two branches 
are always connected with each other, but our second-order approximation of the 
lattice Hamiltonian \eqref{hlat2} describes only their behavior in proximity of the 
Weyl point projections and fails in depicting the global behavior of the Fermi 
arc across the Brillouin zone. The regime $\pi \le \gamma \le 2\pi$ may 
indeed correspond to systems and surfaces in which a Fermi arc connects 
Weyl points in adjacent Brillouin zones (see, for example, the recent experimental results in Ref. \onlinecite{morali2019}).

The above approach is very general and applies to other Hamiltonians in spaces with boundary as well. 
We stress in this respect that once the Hamiltonian $H$ is fixed, there is no further freedom for choosing neither 
the boundary condition nor the number of parameters, describing 
its self-adjoint extensions. In fact, the boundary condition is 
uniquely determined by requiring the Hermiticity (\ref{selfadj}) of $H$, whereas the number of 
parameters is fixed by the indices $n_\pm$ associated to $H$ (see e.g. Ref. \onlinecite{RS} for details). 
If $n_- = n_+\equiv n$ according to the Von Neumann theorem\cite{RS} $H$ admits a $n^2$-parameter 
family self-adjoint extensions. According to (\ref{bc}) a single 
angle $\gamma$ determines all self-adjoint extensions of the Hamiltonian (\ref{ham1}), which implies that in our case $n = 1$.  
This is related to our choice of a minimal 2-component 
model Hamiltonian for the description of the Weyl semimetal. 
We observe, however, that in the literature Dirac Hamiltonians involving 4-component 
spinors are often adopted to describe this kind of systems (see, for example, \cite{faraei18,faraei2019} 
and the analogous case for graphene \cite{akhmerov2008}). In that case the self-adjoint extensions 
depend on a larger set of parameters. For the sake of simplicity we base our analysis on the model in Eq. \eqref{ham}, which allows us 
to capture all the main physical features of the surface transport of Weyl semimetals, 
without resorting to larger spinors.

Summarizing, the Hamiltonian $H_\gamma$ on the half space ${\mathbb R}^3_+$ 
has quite remarkable spectral properties, which represent the core of the quantum field 
description developed below. In particular, the main physical properties of the surface states can be derived from the Hamiltonian \eqref{ham} for boundary conditions with $0<\gamma<\pi$. In the remainder of the paper, 
this regime will be assumed.

\subsection{Quantum field approach}

The strategy is well known and aims at the construction of a quantum field 
\begin{equation} \label{tfield} 
\Psi (t,{\bf r}) = \Psi_{\rm s} (t,{\bf r}) + \Psi_{\rm b} (t,{\bf r})\,,
\end{equation}
where $\Psi_{\rm s}$ and $\Psi_{\rm b}$ collect the bulk and surface contributions. 
The fundamental requirements are that the time evolution of $\Psi$ is 
generated by $H_\gamma $ and that $\Psi$ satisfies 
the canonical equal-time anti-commutation relations. In order to write the solution in explicit form 
we adopt for later convenience a Dirac type formulation and introduce the following notation: 
we label by $a_{\rm s}$ and $a_{\rm s}^\dag$ 
the annihilation and creation operators for surface quasiparticles with $\varepsilon_{\rm s}>0$, and by $b_{\rm s}$ and 
$b_{\rm s}^\dag$ the annihilation and creation operators for surface quasiholes, such that:
\begin{align}
a_{\rm s}(p)&= c_{\rm s}(p) \Theta\left[\varepsilon_{\rm s}(p)\right]\,,
\label{as}\\
b_{\rm s}(-p)&= c_{\rm s}^\dag (p)\Theta\left[-\varepsilon_{\rm s}(p)\right]\, ,
\end{align}
where the operators $c_{\rm s}(p)$ are annihilation operators of the electrons on the surface and 
$\Theta$ is the Heaviside step function. Analogously,  we set in the bulk 
\begin{align}
a_{\rm b}({\bf p})&= c_{\rm b}({\bf p}) \Theta\left[\varepsilon_{\rm b}({\bf p})\right]\,,\\
b_{\rm b}(-{\bf p})&= c_{\rm b}^\dag ({\bf p})\Theta\left[-\varepsilon_{\rm b}({\bf p})\right]\, ,
\label{bb}
\end{align}
where $c_{\rm b}({\bf p})$ are the annihilation operators of the electrons in the bulk. With these conventions 
$\Psi_{\rm s} $ and $\Psi_{\rm b}$ take the form 
\bwt
\begin{align}
\label{s3}
\Psi_{\rm s} (t, \bm r) &= \int\displaylimits \frac{d^2p}{(2 \pi)^2}  \left [ a_{\rm s}(p)  
\zeta_{\rm s}^{(+)}({\bm r},p)  \e^{-i t \left|\varepsilon_{\rm s} (p)\right|} + 
b_{\rm s}^\dag(-p)  \zeta_{\rm s}^{(-)}({\bm r},p)  \e^{i t \left|\varepsilon_{\rm s} (p)\right|}\right ]\, , \\ 
\label{b3}
\Psi_{\rm b} (t, \bm r) &= \int\displaylimits \frac{d^3p}{(2 \pi)^3}\left [  a_{\rm b}(\bm p)  
\zeta_{\rm b}^{(+)}(\bm {r, p})  \e^{-i t \left|\varepsilon_{\rm b} (\bm p)\right|} + b_{\rm b}^\dag(-\bm p)  
\zeta_{\rm b}^{(-)}(\bm {r, p})  \e^{i t \left|\varepsilon_{\rm b} (\bm p)\right|}\right ] \, ,
\end{align}
\ewt 
in terms of the complete system (\ref{a}) of eigenfunctions of $H_\gamma$ and the creation and annihilation 
operators (\ref{as}-\ref{bb}), satisfying the canonical anti-commutation relations. 
We observe that the surface and the bulk components (\ref{s3},\ref{b3}) obey separately the 
equation of motion. This is not he case for the equal-time canonical anti-commutation relations, 
which follow from the completeness of the energy eigenstates and hold therefore only for the total 
field (\ref{tfield}). 

At this point, the choice of representations of the oscillator algebras, generated by (\ref{s3},\ref{b3}),  
is the only freedom we are left with. 
Since our goal is to study the Weyl semimetals at finite temperature and density, we will adopt below the Gibbs 
representation \cite{bratteli}, keeping in general different (inverse) temperatures and chemical potentials 
$\{\beta_{\rm s},\mu_{\rm s}\}$ and $\{\beta_{\rm b},\mu_{\rm b}\}$. In terms of the Fermi distribution 
\begin{equation} 
f(\varepsilon;\beta,\mu) = \frac{\Theta(\varepsilon)}{1+\e^{\beta(\varepsilon - \mu)}}\, , \qquad \beta=1/k_B T\, , 
\label{f}
\end{equation}
the non-vanishing two-point functions are:
\begin{align}
&\left\langle a_{\rm s}^\dag(p) a_{\rm s}(q)\right\rangle = 
f\left(\varepsilon_{\rm s}(p),\beta_{\rm s},\mu_{\rm s} \right)(2\pi)^2 \delta(p-q)\,, 
\label{fermi1}\\
&\left\langle b_{\rm s}^\dag(-p) b_{\rm s}(-q)\right\rangle = f\left(-\varepsilon_{\rm s}(p),\beta_{\rm s},-\mu_{\rm s} \right)(2\pi)^2 \delta(p-q)\,. 
\label{fermi2}
\end{align}
For the bulk excitations one has instead 
\begin{align}
&\left\langle a_{\rm b}^\dag(\bm p) a_{\rm b}(\bm q)\right\rangle = 
f\left(\varepsilon_{\rm b}(\bm p),\beta_{\rm b},\mu_{\rm b} \right)(2\pi)^3 \delta(\bm {p-q})\,, \label{fermi3}\\
&\left\langle b_{\rm b}^\dag(-\bm p) b_{\rm b}(-\bm q)\right\rangle = 
f\left(-\varepsilon_{\rm b}(\bm p),\beta_{\rm b},-\mu_{\rm b} \right)(2\pi)^3 \delta(\bm {p-q})\,.  
\label{fermi4}
\end{align} 
We observe that there is no interference between surface and bulk oscillators and that 
all higher point correlation functions can be expressed in terms of (\ref{fermi1}-\ref{fermi4}). 

At this stage we can construct and investigate the basic physical observables of the system. 
Let us consider for instance the particle density operator 
\begin{align} 
:\Psi^\dagger \Psi:(t,{\bf r}) =\, :\Psi_{\rm s}^\dagger \Psi_{\rm s}:(t,{\bf r}) + :\Psi_{\rm b}^\dagger \Psi_{\rm b}:(t,{\bf r}) \; \; 
\nonumber \\
+ :\Psi_{\rm s}^\dagger \Psi_{\rm b}:(t,{\bf r})+:\Psi_{\rm b}^\dagger \Psi_{\rm s}:(t,{\bf r})\, , 
\label{d1}
\end{align}
where $:\cdots :$ stands for the normal ordering with respect to the creation and annihilation operators (\ref{as}-\ref{bb}). 
Since the surface and the bulk operators have vanishing mixed two-point functions, one finds  
\be 
\langle :\Psi^\dagger \Psi:(t,{\bf r})\rangle  =\, \langle :\Psi_{\rm s}^\dagger \Psi_{\rm s}:(t,{\bf r})\rangle  + 
\langle :\Psi_{\rm b}^\dagger \Psi_{\rm b}:(t,{\bf r})\rangle \, , 
\label{d2}
\ee
which implies that at the level of mean values there is no interplay between surface and bulk degrees 
of freedom. This is a general feature, which allows us to treat separately the mean values of the 
surface and bulk currents as well.  

Let us stress finally that the above quantum field theory setting works directly in the thermodynamic limit of our system.

\section{Surface currents and Hall conductivity} 
\label{zeroT}

\subsection{Mean value of the surface current}

The anomalous Hall conductivity of Weyl semimetals is determined by their Fermi arcs. In the minimal model \eqref{hlat2}, a simple decomposition of the 3D Hamiltonian \eqref{ham} into a two-dimensional set of systems parameterized by $p_x$ shows that the number of surface states defining the the Fermi arc is proportional to the distance $2p_0$ between the projections of the Weyl points in the surface Brillouin zone. In particular, each 2D system defined by $H_{p_x}(p_y,p_z)$ constitutes a Chern insulator with chiral gapless edge modes for $-p_0<p_x<p_0$, and this implies that the contribution to the anomalous Hall conductivity of the surface states in the ballistic regime at zero temperature and half filling is given by $\sigma_H = e^2p_0/\pi h$ \cite{balents11,yang2011}. This value of the anomalous Hall conductivity is universal and does not depend on the boundary conditions of the system.
Other physical quantities, as, for example, the behavior of $\sigma_H$ at finite chemical potential, depend instead on the boundary condition \eqref{bc}. In the following, we analytically study the physics of the surface modes focusing on several characteristics which are determined by the value black $0< \gamma < \pi$ of the surface polarization.

To investigate the transport properties of the system generated by the surface states we consider 
the surface current ${\bf j}(t,{\bf r})$, satisfying the continuity equation 
\be 
\partial_t n_{\rm s}(t,{\bm r}) = -\bm \nabla {\bf j} (t,{\bm r})\, ,  
\quad n_{\rm s}(t,{\bm r})= \, :\Psi_{\rm s}^\dag \Psi_{\rm s}: (t,{\bm r})\, .  
\label{js}
\ee
Using the equations of motion, one finds for ${\bf j}(t,{\bf r})$: 
\begin{align}
j_x(t,{\bm r}) &= \frac{iv }{2p_0} :\left [ \left (\der_x \Psi_{\rm s}^\dagger \right ) \sigma_x \Psi_{\rm s}  - 
\Psi_{\rm s}^\dagger \sigma_x \left (\der_x \Psi_{\rm s} \right ) \right ]:(t, \bm r )\, ,  \\
j_y (t,\bm r) &= v : \Psi_{\rm s}^\dagger \sigma_y \Psi_{\rm s}: (t,\bm r)\, ,  \label{jyop} \\
j_z (t,\bm r) &= v : \Psi_{\rm s}^\dagger \sigma_z \Psi_{\rm s} :(t,\bm r)\, .
\end{align}
Now, adopting the two-point functions (\ref{fermi1},\ref{fermi2}) and the explicit form (\ref{A1}) of the surface eigenfunctions one obtains 
\bwt
\begin{align}
\langle n_{\rm s} (t, \bm r ) \rangle  &=  2\int \frac{d^2p}{(2 \pi)^2}   
\Theta \left[\tilde{p}_z(p)\right ]\tilde{p}_z(p) \e^{-2 \tilde{p}_z(p) z }  
\left\{ f\left (\varepsilon_{\rm s}(p);\beta_{\rm s},\mu_{\rm s}\right ) - f\left (-\varepsilon_{\rm s}(p);\beta_{\rm s},-\mu_{\rm s}\right )\right\}  \,, \label{rhoT}\\
\left\langle j_x (t,\bm r)\right\rangle &= \left\langle j_z (t,\bm r)\right\rangle =0\,,\\
\left\langle j_y (t,\bm r)\right\rangle &= 2 v\sin \gamma \int \frac{d^2p}{(2 \pi)^2}  
\Theta \left[\tilde{p}_z(p)\right ]\tilde{p}_z(p) \e^{-2 \tilde{p}_z(p) z }  
\left\{ f\left (\varepsilon_{\rm s}(p);\beta_{\rm s},\mu_{\rm s}\right ) - f\left (-\varepsilon_{\rm s}(p);\beta_{\rm s},-\mu_{\rm s}\right )\right\}  \,, \label{jyexp}
\end{align}
\ewt
with $\tilde{p}_z(p)$ given by (\ref{g}). 
As expected, the mean values (\ref{rhoT}-\ref{jyexp}) are time independent (invariance under time translations) and 
$(x,y)$-independent (invariance under space translations in the $(x,y)$-plane). 
Moreover, they manifestly satisfy the continuity equation (\ref{js}) and the relation  
\begin{equation} \label{jy}
\left\langle j_y (t, \bm r)\right\rangle = v \sin \gamma \langle n_{\rm s} (t,\bm r ) \rangle\,.
\end{equation}

Equations (\ref{rhoT}-\ref{jyexp}) provide the distribution of the density of the electrons and 
the current generated by the surface states as a function of the distance $z$ from the boundary.  We point out that the particle density in Eq. \eqref{rhoT}
is a different quantity compared to the density of the surface states, which, instead, can be easily derived from Eq. \eqref{s2} (see Section \ref{num}).
The surface current is dissipationless in our model, since we neglect scattering effects from surface 
to bulk states, and it is responsible for the anomalous quantum Hall conductivity. Its 
dependence from $x$ and $y$ is trivial due to the translational invariance in these directions. 
The surface current along the $\hat{x}$ direction has a vanishing expectation value due to the 
contributions of states with positive and negative $p_x$ canceling each other.

\subsection{The anomalous Hall conductivity}

In the ballistic regime, the surface currents are dissipationless; this implies that, in a typical 2-terminal transport measurement with two external leads attached to a Weyl semimetal scatterer, the surface states can be considered in equilibrium with the leads they originate from. This is analogous to the standard quantum Hall devices in two dimensions and it allows us to consider the distribution of the surface states at a fixed chemical potential. The Hall conductivity can thus be obtained from the derivative of the expectation value of the current with respect to the chemical potential inherited by the source lead.

In order to investigate the anomalous Hall conductivity it is instructive to consider the $p_x$ components $\hat{j}_y(p_x,z,\mu_s)$ of the surface current density such that $j_y(z,\mu_s) = \int dp_x \hat{j}_y(p_x,z,\mu_s)$. In particular, we can define the local contribution of each surface state labelled by $p_x$ to the total differential conductivity $\sigma_H$:
\begin{equation} \label{sigmahat}
\hat{\sigma}(p_x,z,\mu_s) = e^2 \partial_{\mu_s} \left\langle \hat{j}_y(p_x,z,\mu_s) \right\rangle , 
\end{equation}
such that:
\begin{equation} \label{anomalcond}
\sigma_H(\mu_s) = \int dz \int dp_x \, \hat{\sigma}(p_x,z,\mu_s)\,.
\end{equation}
The momentum integrals in Eqs. \eqref{jyexp} and \eqref{sigmahat} cannot be expressed in simple closed form in the general case. However, we can invert the order of the integrations, and determine $\sigma_H(\mu_s)$ by integrating first the space coordinate $z$, and then the momenta. This allows for a general expression of the Hall conductivity as a function of the chemical potential $\mu_s$, the inverse temperature $\beta$ and the boundary polarization $\gamma$:
\begin{multline} \label{Hallgeneral}
\sigma_H(\beta,p_0,\mu_s,\gamma) =\\ 
= -\frac{e^2}{2\pi^2\hbar} \sqrt{\frac{p_0 \pi |\cos(\gamma)|}{2v\beta_s}}\, 
{\rm Li}_{\frac{1}{2}}\left [-\e^{\frac{\beta(2\mu_s \cos(\gamma)+vp_0)}{2|\cos(\gamma)|}}\right ],
\end{multline}
where we reintroduced the Planck constant $\hbar$ for clarity. Here ${\rm Li}_{1/2}$ is a polylogarithm function \cite{mathbooks}.

\begin{figure}[tb]
\includegraphics[width=\columnwidth]{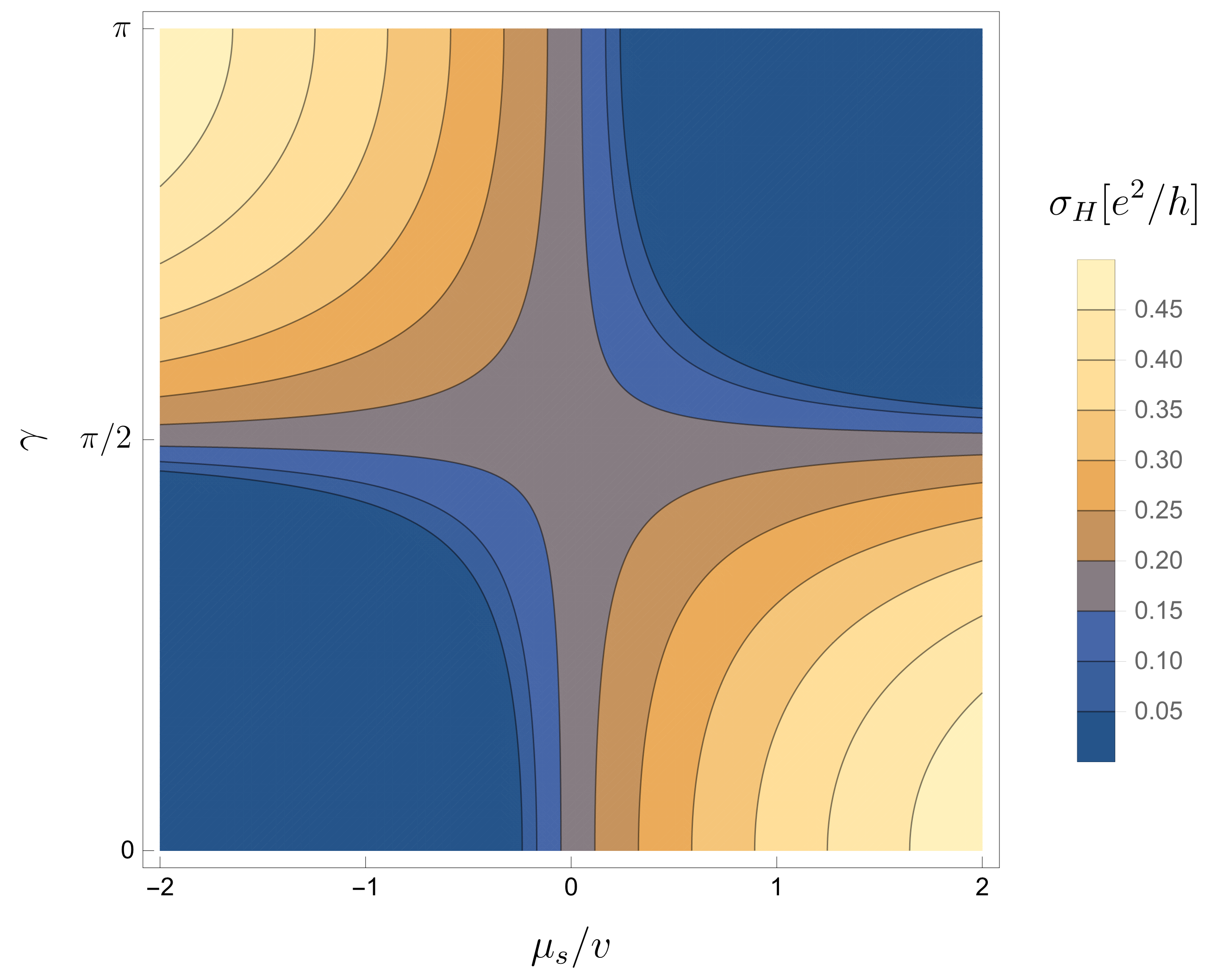}\llap{
  \parbox[b]{16.5cm}{(a)\\\rule{0ex}{6.5cm}}}
\includegraphics[width=\columnwidth]{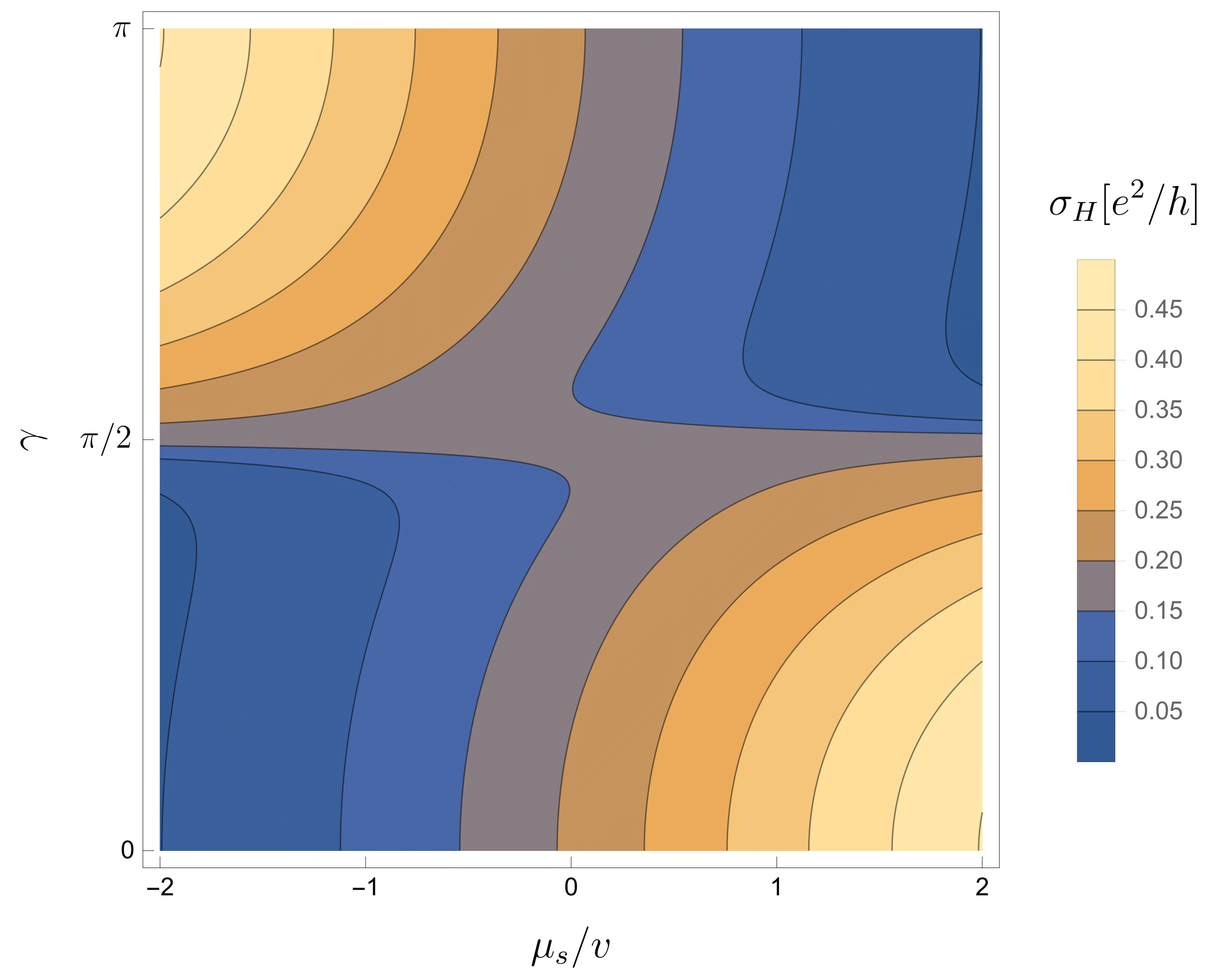}\llap{
  \parbox[b]{16.5cm}{(b)\\\rule{0ex}{6.5cm}}}
	
\caption{Anomalous Hall conductivity $\sigma_H$ as a function of the chemical potential $\mu_s/v$ and the boundary polarization $\gamma$ for $p_0=\pi/6$. (a) Zero temperature case defined by Eq. \eqref{hall};  for $\mu_s=0$ or $\gamma=\pi/2$, $\sigma_H$ assumes the universal value $e^2p_0/\pi h =(1/6) e^2/h$. (b) Conductivity at temperature $k_BT_s/v=\pi/3$ calculated from Eq. \eqref{Hallgeneral}; for $\gamma=\pi/2$, $\sigma_H$ assumes the universal value $e^2p_0/\pi h =(1/6) e^2/h$ also at finite temperature.}
\label{fig:sigmaHt}
\end{figure}

In the zero-temperature limit, $\beta_s \to \infty$, the anomalous Hall conductance \eqref{Hallgeneral} becomes:
\begin{equation} 
\label{hall}
\sigma_H (T=0,p_0,\mu_s,\gamma) =  e^2 \frac{\sqrt{p_0 [p_0 + 2(\mu_s/v) \cos (\gamma)]}}{2\pi^2 \hbar}\,.
\end{equation}
This expression is proportional to the density of surface states; \textcolor{black}{it depends in general on the specific choice of the function $g(p_x)$ and, in our case,} it is valid for $vp_0 + 2\mu_s \cos (\gamma) > 0$. For values of $\mu_s$ outside this regime at $T_s=0$, the differential anomalous Hall conductivity vanishes because the surface states are either completely empty (for $\cos \gamma >0$) or completely filled (for $\cos \gamma <0$); the dispersion relation \eqref{s2} has indeed a minimum or a maximum for $\cos\gamma \gtrless 0$ respectively due to the constraint $\tilde{p}_z(p)>0$.
Eq. \eqref{hall} provides the known result $\sigma_H=e^2p_0/\pi h$ for $\mu_s \to 0$ and arbitrary boundary conditions $0<\gamma<\pi$, such that this value is universal (in the ballistic regime). Furthermore, we observe that for $\gamma=\pi/2$, the conductivity $\sigma_H$ is independent of $\mu_s$; for $0<\gamma<\pi/2$, it increases with $\mu_s$, whereas it decreases for $\pi/2 < \gamma < \pi$ [see Fig. \ref{fig:sigmaHt} (a)]. This shows the importance of the boundary conditions in the determination of the non-universal corrections to the anomalous Hall conductivity.

\begin{figure}[tb]
\includegraphics[width=\columnwidth]{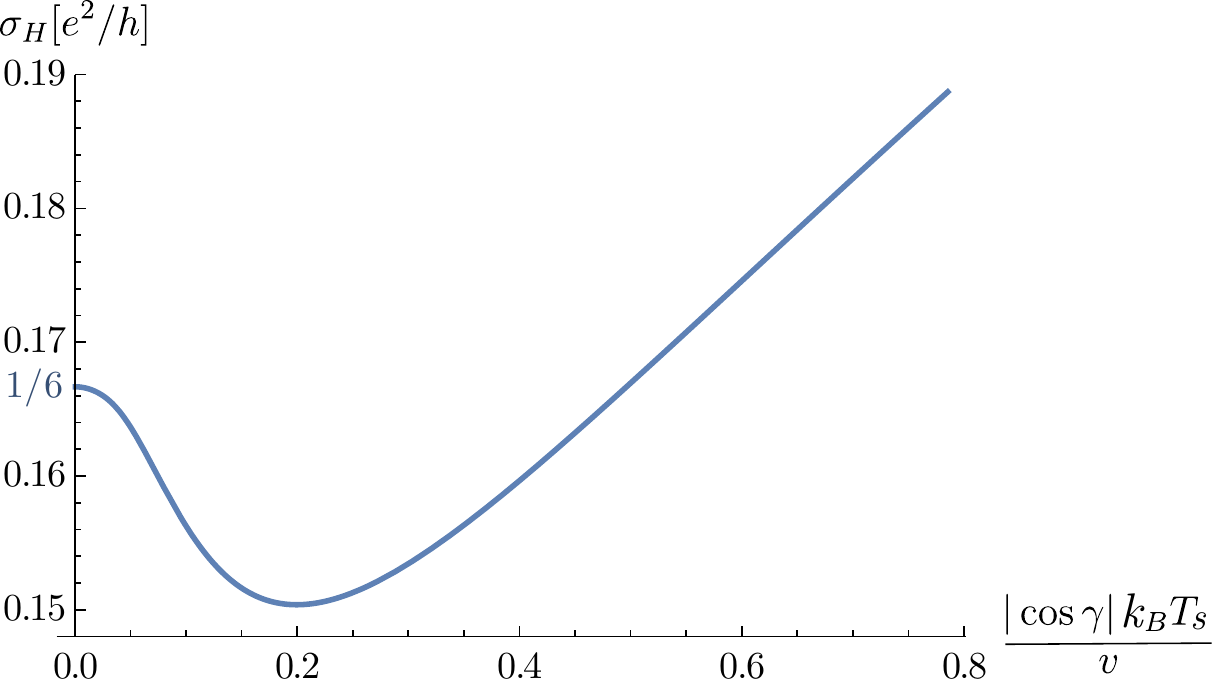}
\caption{Anomalous Hall conductivity $\sigma_H$ for a system with $p_0=\pi/6$ and $\mu_s=0$, calculated with Eq. \eqref{Hallgeneral} as a function of the parameter $|\cos\gamma|k_BT_s/v$. For $T=0$ or $\gamma=\pi/2$, the universal Hall conductivity $e^2p_0/\pi h =(1/6) e^2/h$ is retrieved. The behavior of $\sigma_H$ is, in general, non-monotonic in $T$.}
\label{fig:sigmamu0}
\end{figure}

For finite temperature the system acquires a non-zero surface conductance also for values of the chemical potential such that $vp_0 + 2\mu_s \cos(\gamma)\le 0$ [see Fig. \ref{fig:sigmaHt} (b)]. The temperature dependence of the Hall conductivity \eqref{Hallgeneral} is non-trivial. For $\mu_s=0$, the conductivity is a function of the rescaled temperature $k_BT_s|\cos\gamma|/v$ and the parameter $p_0$. In particular, $\sigma_H$ is non monotonic in $T_s$ (see Fig. \ref{fig:sigmamu0}): by increasing $T_s$ from zero, the conductivity decreases from the universal value $e^2p_0/\pi h$ to a minimum which depends on $p_0$. Then it increases again and, asymptotically, it grows proportionally to $\sqrt{p_0k_BT_s|\cos\gamma|/v}$. This behavior reflects the particular density of states of the surface modes which is determined by 
the dispersion (\ref{s2}) combined with the constraint $\tilde{p}_z(p)>0$.

\subsection{The case of straight Fermi arc $\gamma=\pi/2$}

For the particular case of straight Fermi arcs, $\gamma=\pi/2$, the anomalous Hall conductivity acquires its universal value $\sigma_H=e^2p_0/\pi h$ and it is independent of both $T_s$ and $\mu_s$, as it can be verified by the limit of Eq. \eqref{Hallgeneral} [see, for instance, Fig. \ref{fig:sigmaHt} (b)].

Besides the total Hall conductivity, the definition of the current density \eqref{jyexp} allows us also to define the current as a function of the distance $z$ from the surface, thus the local contribution \eqref{sigmahat} to the Hall conductivity. 
To analyze the local transport properties of the system we begin by investigating the case of a straight Fermi arc, $\gamma=\pi/2$, for which $\varepsilon_{\rm s}$ depends on $p_y$ only. For this boundary conditions $\left\langle j_y(z)\right\rangle$ becomes independent of the chemical potential and temperature:
\begin{equation}
\left\langle j_y (z)\right\rangle = \frac{\mu_s} {4 \pi^2 z^2} \sqrt {p_0 z} \left [(1+2p_0 z)\, D_+(\sqrt {p_0 z}) - 
\sqrt {p_0 z}\, \right]; \label{jyav}
\end{equation}
hereafter the function $D_+$ labels the Dawson integral:
\begin{equation} \label{dawson}
D_+(\xi)=\e^{-\xi^2} \int_0^\xi \e^{\eta^2} d\eta\,.
\end{equation}

\begin{figure}[tb]
\includegraphics[width=\columnwidth]{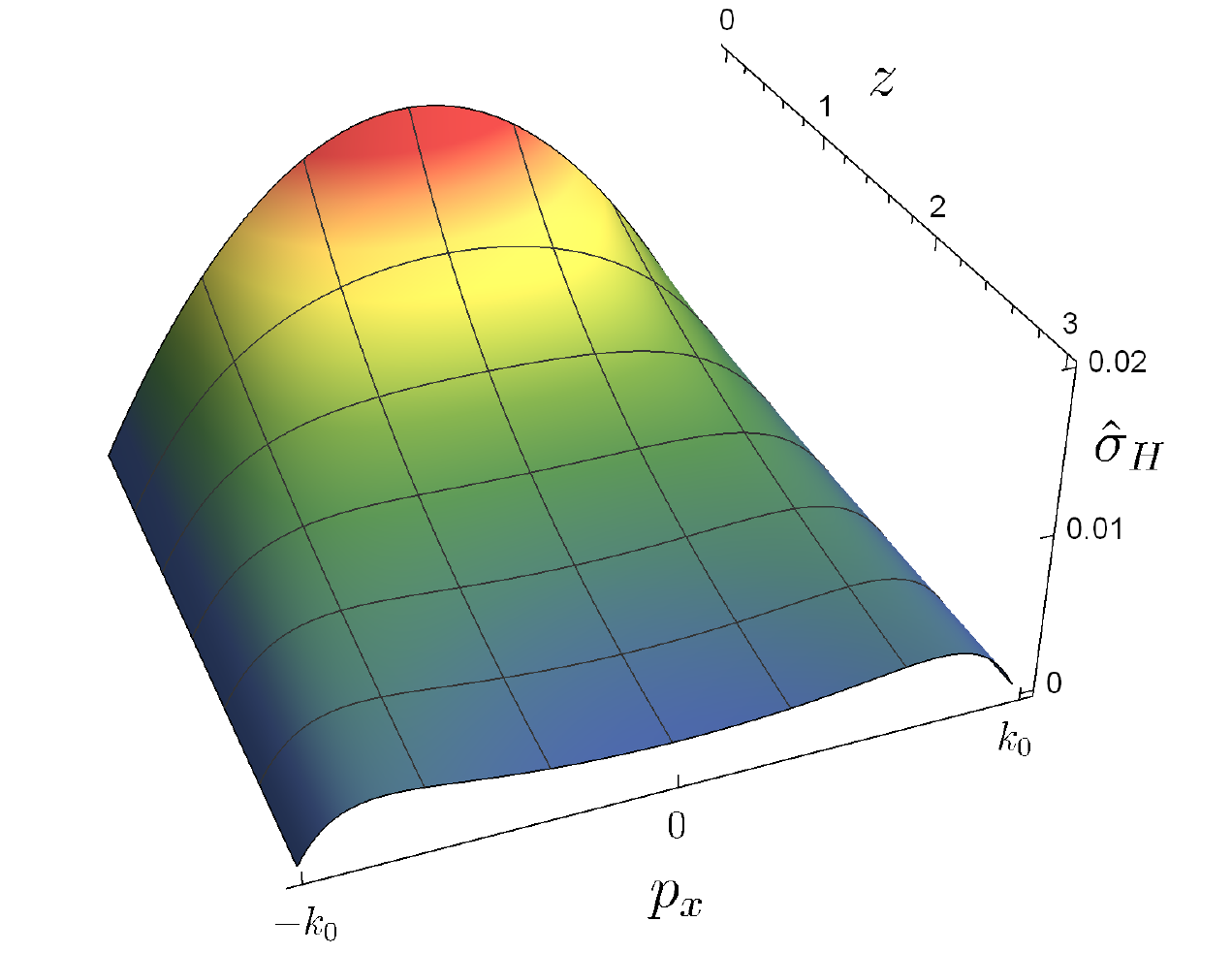}
\caption{Density of the anomalous Hall conductance in Eq. \eqref{sigmah2} for $\gamma=\pi/2$ and $p_0=\pi/5$ as a function of the distance $z$ from the surface and the momentum $p_x$ along the Fermi arc.} \label{fig:sigmah2}
\end{figure}

The current \eqref{jyav} at $\gamma=\pi/2$ is consistent with the following local conductivity of the surface state labelled by $p_x$:
\begin{equation} \label{sigmah2}
\hat{\sigma}(p_x,z) = \frac{e^2}{h} \frac{(p_0^2-p_x^2) \e^{-z(p_0^2-p_x^2)/p_0}}{2 \pi p_0}\, ,
\end{equation}
independent of $\mu$. We plot $\hat{\sigma}(p_x,z)$ for $\gamma=\pi/2$ in Fig. \ref{fig:sigmah2}; its decay length in the bulk is given by $g(p_x)^{-1}$: as expected, the states of the Fermi arc with $p_x$ approaching the Weyl point projections at $p_x=\pm p_0$ progressively penetrate deeper in the bulk and their contribution to the conductivity is weaker for $z=0$ but decays slower with the distance. For $p_x$ in the center of the Fermi arc, instead, the surface states are more localized and their contribution is stronger for $z=0$.

\subsection{The zero-temperature limit}

For generic values of the boundary condition parameter, the integrals in Eqs. \eqref{rhoT} and \eqref{jyexp} cannot be written in simple closed forms. The results simplify in the zero-temperature limit, that allows us to clearly evaluate the role of the boundary conditions and the chemical potential. For $T\to 0$ we can substitute the Fermi function (\ref{f}) with:
\begin{equation}
f(\varepsilon;\beta,\mu)\stackrel{T\to0}{\longrightarrow} \Theta\left[\varepsilon\right]\Theta\left[\mu - \varepsilon\right]
\end{equation}
Concerning the operators $n_{\rm s}$ and $j_y$ in Eqs. \eqref{rhoT} and \eqref{jyexp}, in the limit $\mu_s \to 0$, the difference of the $T_s \to 0$ limits of the Fermi distributions in the integral is meant to return the contribution of the zero-energy Fermi arc states.
From Eq. \eqref{jy} we observe that the angle $\gamma$ fixes the direction of the surface current density in our model, which gets inverted for $\pi < \gamma < 2\pi$. This inversion can be obtained in the physical systems only for Fermi arcs that connect the two projections of the Weyl points by winding across the Brillouin zone: our approximation of the lattice Hamiltonian \eqref{hlat2}, however, is accurate only for momenta relatively close to the Weyl points (and small values of $p_0$); therefore, Eq. \eqref{ham} does not describe correctly the physics of the whole surface Brillouin zone, resulting in unbounded Fermi arcs for $\pi < \gamma < 2\pi$.

The momentum integral in Eq. \eqref{rhoT} determines the decay of the total density of the surface states in the bulk; in particular, for large values of $z$ we may approximate the density with the asymptotic expansion:
\begin{equation} \label{surfdensity}
\left\langle n_{\rm s}(\bm r)\right\rangle \approx \frac{\sqrt{p_0\left(p_0 + 2(\mu_s/v) \cos \gamma\right)}-p_0}{4\pi^2 z^2 \cos\gamma } \,.
\end{equation}
This relation is defined only for $p_0 + 2(\mu_s/v) \cos \gamma >0$. For values of $\mu_s$ outside this range of validity, the surface states are either completely filled (for $\cos \gamma <0$) or completely empty (for $\cos \gamma >0$) and $\left\langle n_{\rm s}\right\rangle \approx -p_0/\left(4\pi^2 z^2\cos\gamma\right)$.
The relation \eqref{surfdensity} implies that the expectation value of the density and current density of the surface state decays as $z^{-2}$ in the bulk. As emphasized by several works \cite{gorbar16,resta18}, this may yield in turn relevant scattering processes between surface and bulk states.
The limit $z\to 0$, instead, provides an estimate of the surface density exactly at the boundary and it is given by:
\begin{equation}
\left\langle n_{\rm s} (z=0)\right\rangle = \frac{\sqrt{p_0}\left(p_0 + 2(\mu_s/v) \cos \gamma\right)^{5/2}-p_0^3}{15\pi^2\cos\gamma\sin^2\gamma}\,.
\end{equation}

\begin{figure}[t]
\includegraphics[width=\columnwidth]{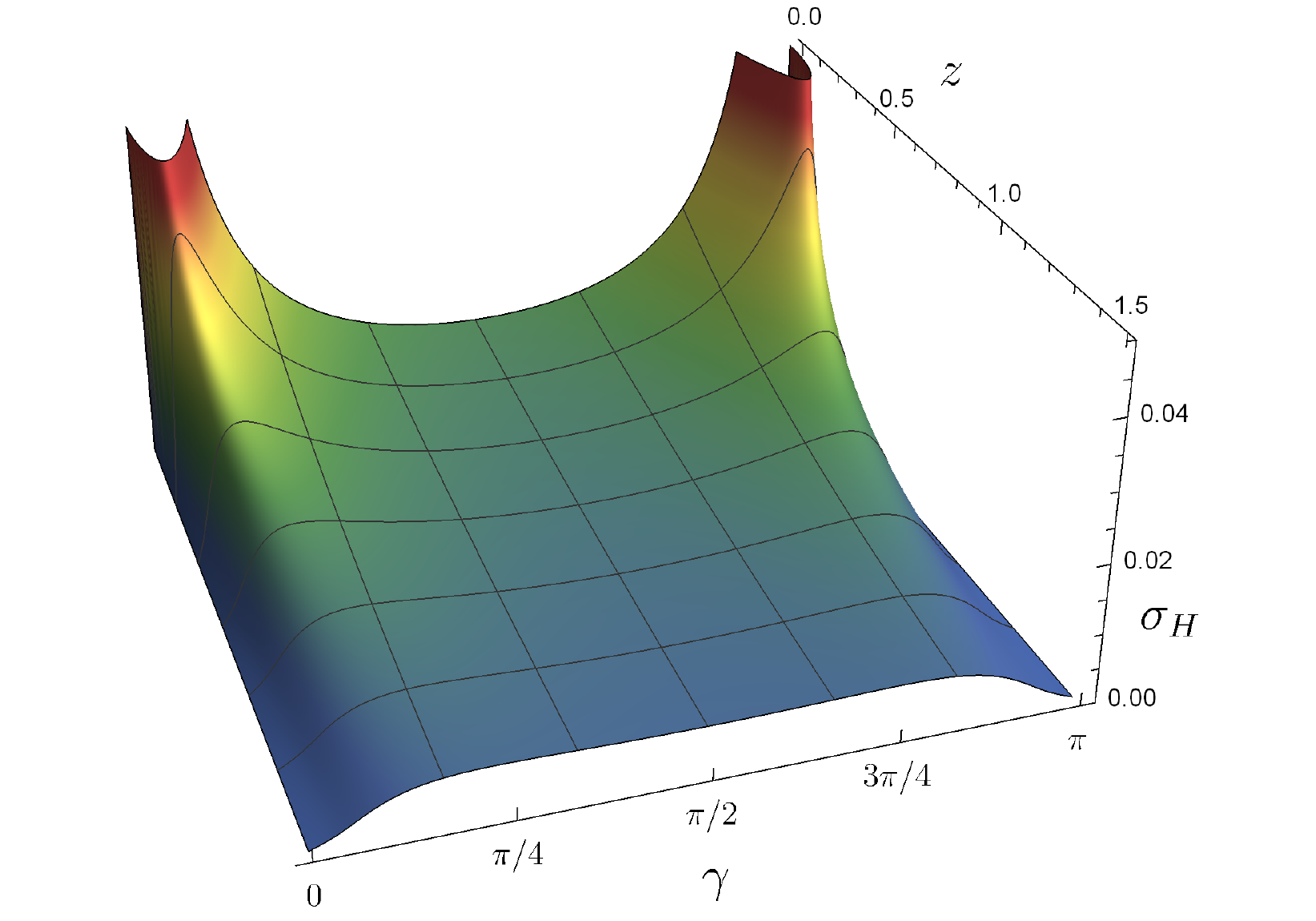}
\caption{Density of the anomalous Hall conductance in Eq. \eqref{conddens} for $\mu_s=0$ and $p_0=\pi/5$ as a function of the distance $z$ from the surface and the boundary condition parameter $\gamma$.} \label{fig:sigmah}
\end{figure}

After defining $\hat{\sigma}(p_x,z,\mu_s)$ from Eq. \eqref{sigmahat}, the integral in the $p_x$ momenta in Eq. \eqref{anomalcond} returns:
\begin{multline} \label{conddens}
\sigma_H(z,\mu_s) = \int dp_x  \hat{\sigma} (p_x) = \frac{e^2}{h}
   \frac{\sqrt {{p_0 z\sin(\gamma)}}} {2 \pi z^2}  \times \\
 \left \{\left [1+2{\tilde p_0} z\right ]\, D_+\left (\sqrt {{\tilde p_0} z}\right ) - 
\sqrt {{\tilde p_0} z}\, \right\},
\end{multline} 
where
\begin{equation}
{\tilde p_0} = \frac{p_0 +2(\mu_s/v) \cos(\gamma)}{\sin(\gamma )} \, .
\end{equation} 
The conductance density \eqref{conddens} is represented in Fig. \ref{fig:sigmah} as a function of the boundary condition parameter $\gamma$ and the distance from the surface $z$. It decays asymptotically as: 
\begin{equation}
\sigma_H(z,\mu_s) \approx \frac{e^2}{h}\frac{\sin \gamma}{2\pi z^2} \sqrt{  \frac{p_0}{p_0 +2(\mu_s/v) \cos(\gamma)}}\,.
\end{equation}

All the previous results are based on the specific model \eqref{ham} which is characterized by a single pair of Weyl cones and can be considered a good approximation for experimental systems with two Weyl cones only \cite{soh2019}. In the case of materials with more Weyl points, however, we expect that each Fermi arc will provide a contribution to the anomalous Hall conductance given by the previous equations (\ref{conddens}-\ref{sigmah2}). The additivity of these contributions can be verified in the limit of small $p_0$, in which several pairs of Weyl points are sufficiently distant from each other in momentum space.

\section{Bulk conductance}
\label{bulkcond}

The total conductance of a Weyl semimetal is given by the sum of its surface and bulk contribution.
In the previous section, we evaluated the contribution of the conductivity due to the surface modes. The bulk conductance, instead, can be estimated through the Landauer approach by considering a Weyl semimetal connected to two external leads and evaluating the transmission amplitude associated to each of their modes. This approach has been applied in Ref. \onlinecite{baireuther14} for the case of a pair of overlapping Weyl cones at the phase transition between topological and normal insulators and further detail can be found in Ref. \onlinecite{beenakker2006} for the two-dimensional case of graphene.

Here we apply the same approach to approximate the conductance of the three-dimensional Weyl semimetal described by the Hamiltonian \eqref{ham}. To this purpose, we consider a system divided into three regions in the $\hat{y}$ directions, with width $W$ in the $\hat{x}$ and $\hat{z}$ directions. In order to estimate the bulk conductance, we consider periodic boundary conditions in the $\hat{x}$ and $\hat{z}$ directions, such that the momenta $p_x$ and $p_z$ are quantized in units of $2\pi/W$. The regions at $y<0$ and $y>L$ constitute two infinite leads, and we describe them with the Hamiltonian \eqref{ham} and a chemical potential $\mu_{\rm lead} \to \infty$, in such a way that the two leads are effectively in a metallic phase with a large density of states. The central region $0\le y \le L$, instead, is characterized by $\mu=0$ and models a Weyl semimetal with chemical potential lying at the level of the two Weyl points.

We impose the continuity of the wavefunction at the interfaces $y=0$ and $y=L$ such that, for each value of $p_x$ and $p_z$ we obtain the transmission probability (see Appendix \ref{app:conductance} for more detail):
\begin{equation} \label{transmission}
\mathcal{T}(p_x,p_z)= \frac{1}{\cosh^2\left[L \sqrt{g(p_x)^2+p_z^2}\right]}\,,
\end{equation}
which generalizes in a straightforward way the result in Ref. \onlinecite{baireuther14} \textcolor{black}{for an arbitrary $g(p_x)$. For our specific choice of $g(p_x)$ and} periodic boundary conditions along the $\hat{x}$ and $\hat{z}$ directions, the bulk conductance at vanishing chemical potential results:
\bwt
\begin{equation} 
\label{transsum}
G_{\rm b} = \frac{e^2}{h} \sum_{p_x,p_z}  \mathcal{T}(p_x,p_z)= 
 \frac{e^2}{h} \sum_{n_x,n_z} \cosh^{-2}\left[2\pi L \sqrt{\frac{\pi^2}{p_0^2}\left(\frac{n_x}{W}+\frac{p_0}{2\pi}\right)^2\left(\frac{n_x}{W}-\frac{p_0}{2\pi}\right)^2+\frac{n_z^2}{W^2}}\right]\,;
\end{equation}
\ewt
where we adopted the notation $p_{x,z} = 2\pi n_{x,z}/W$. We observe that, in general, $G_{\rm b}$ is not scale invariant due to the parameter $p_0$, and it does not depend only on $L/W$. To evaluate the general behavior of the conductance $G_{\rm b}$ it is useful to distinguish two regimes: a fine-tuned regime, in which $p_0$ is an integer multiple of $2\pi/W$, and a standard regime in which $p_0$ is not a multiple of $2\pi/W$.

The fine-tuned regime is special because the Weyl points lie exactly on one of the momenta of the Brillouin zone, therefore there are two bulk zero-energy modes that contribute with a quantum of conductance to the bulk transport, independently on $L$, and the conductance decreases asymptotically to $2e^2/h$ for $L/W\to \infty$. This case is analogous to the result of Ref. \onlinecite{baireuther14} for periodic boundary conditions, and the conductance of the Dirac semimetal in \onlinecite{baireuther14} is recovered in the limit of large $p_0L$, where the contribution of the bulk modes to $G_{\rm b}$ is compatible with having two well-separated Weyl cones. 

The most realistic scenario is the one with $p_0 \neq n_0 2\pi/W$ ($n_0 \in \mathbb{N}$). Differently from the fine-tuned regime, the system has a vanishing conductance in the limit $L/W\to \infty$ for $p_0L \gg 1$. In order to estimate the conductance in this case, we may consider the behavior of the transmission probability in proximity of the two Weyl points, where $\mathcal{T}$ is maximized. Let us consider the case $n_x \approx 2\pi p_0 W$. The maximum value of $\mathcal{T}$ can be approximated by observing that:
\[
\min_{n_x}\left[\left(\frac{n_x}{W}+\frac{p_0}{2\pi}\right)^2\left(\frac{n_x}{W}-\frac{p_0}{2\pi}\right)^2\right] \lesssim \left(\frac{p_0}{\pi}\right)^2\frac{1}{W^2}
\]
such that $\mathcal{T}_{max} \approx \cosh^{-2}\left(2\pi {L}/{W}\right)$. By considering the decay of $\mathcal{T}$ with $n_x$ and $n_z$ away from the Weyl points, we conclude that the system has a vanishing conductance for $L/W\to \infty$.

Let us finally address the limit of close Weyl points, that describes a system approaching a phase transition in which the topological semimetal phase may be gapped. We model this regime by considering $p_0 W \ll 1 $ and we consider a wire geometry, thus $L \gg W$. In this situation we can estimate the behavior of the conductance in the following limits: for $ p_0 L \ll 2$, the conductance goes to $G_{\rm b} \to e^2/h$ because of the contribution of the term $n_x=n_z =0$; for $ p_0 L \to \infty$, instead, the conductance vanishes.

For a system with surfaces at $z=1$ and $z=W$, we assume that the surface at $z=W$ displays an opposite polarization with respect to the one in $z=1$ (as suggested by our numerical results based on the model \eqref{hlat2}, see also Ref. \onlinecite{beenakker18}). In this case:
\begin{equation}
\left ( \sigma_x \cos \gamma + \sigma_y \sin \gamma \right )  \psi (\bm r )  \,  \bigr |_{z=W}  =  -\psi (\bm r )  \,  \bigr |_{z =W} \, . 
\label{bcopposite}
\end{equation}
By considering the Hamiltonian \eqref{ham}, this implies that the values of $p_z$ must be taken as:
\begin{equation}
p_z = \left(n_z + \frac{1}{2} \right) \frac{\pi}{W-1}\,, \quad \text{with } n_z\ge 0\,,
\end{equation} 
analogously with the two-dimensional case of graphene \cite{beenakker2006}.  This quantization of the momenta must be considered in calculating the bulk conductance $G_{\rm b}$ and the Weyl semimetal acquires a total conductance of the form:
\begin{equation}
G \approx G_{\rm b} + \sigma_H W  = G_{\rm b} + e^2p_0W/h\,.
\label{twocond}
\end{equation}
We conclude that the total conductance indirectly depends on the boundary conditions via the quantization of the momenta orthogonal to the surfaces: \textcolor{black}{The bulk term in $G$ is indeed non-universal and depends, in general, on the choice of $g(p_x)$ and the boundary conditions, whereas the surface term represents the universal anomalous Hall conductance for vanishing chemical potential and temperature. }
We finally observe  that, for more general boundary conditions with independent polarizations $\gamma$ and $\gamma'$ on the surfaces at $z=1$ and $z=W$, the determination of the correct set of momenta $p_z$ gives rise, in general, to non-analytical solutions. 

Eq. \eqref{transmission} describes the transmission probability in the system \eqref{ham} with two Weyl points; we observe, however, that it can be extended also to materials characterized by well-separated dipoles of Weyl points, a common experimental situation \cite{hasan17,yan17}. Indeed, the transmission coefficient $\mathcal{T}(p_x, p_z)$ decays exponentially with the distance in momentum space from the pair of Weyl points, hence the value of $G_{\rm b}$ is dominated by the states in proximity of the Weyl pair. Therefore we expect that in a material with small $p_0$ and pairs of Weyl points sufficiently far from each other, the contribution of each pair to the bulk conductance will approximately add to each other. In this case, the resulting conductance $G_{\rm b}$ can be approximated with the sum over all the Weyl dipoles of the value \eqref{transsum} of a single Weyl pair.

\section{Numerical comparisons}
\label{num} 
In this section we verify numerically the analytical results of the previous sections. For this task, we use the {\it Kwant} code \cite{kwant} to simulate the following Hamiltonian of spin-1/2 fermions on the cubic lattice, corresponding to Eq. \eqref{hlat2}:
\begin{multline}
H_{\rm lat} = -\frac{\tilde{v}}{2} \sum_{{\bf r}} \left(c^{\dag}_{{\bf r} + \hat{x}} \sigma_x c_{\bf r} + {\rm H. c.}\right) + b\sum_{{\bf r}}c^{\dag}_{{\bf r}} \sigma_x c_{\bf r} \\
- \mu \sum_{{\bf r}}c^{\dag}_{{\bf r}}  c_{\bf r}  +\frac{v}{2} \sum_{{\bf r}} \left( c^{\dag}_{{\bf r} + \hat{y}} \left[i\sigma_y-\sigma_x\right] c_{\bf r}+ {\rm H. c.}\right) \\
+ \frac{v}{2} \sum_{{\bf r}} \left( c^{\dag}_{{\bf r} + \hat{z}} \left[i\sigma_z -\sigma_x \right] c_{\bf r}+ {\rm H. c.}\right) \, ,
\label{latHam}
\end{multline}
with $b=v\left(2+\cot p_0\right)$.
In the following, we will focus on finite size systems where the Weyl semimetal constitutes a "scattering region" with size $W_x \times L \times W_z$ (in units of the lattice spacing $a \equiv 1$), and we will adopt different boundary conditions. 

\begin{figure*}[t]
\begin{flushleft}
\vspace{0.01\textwidth}
\includegraphics[height=0.35\textwidth]{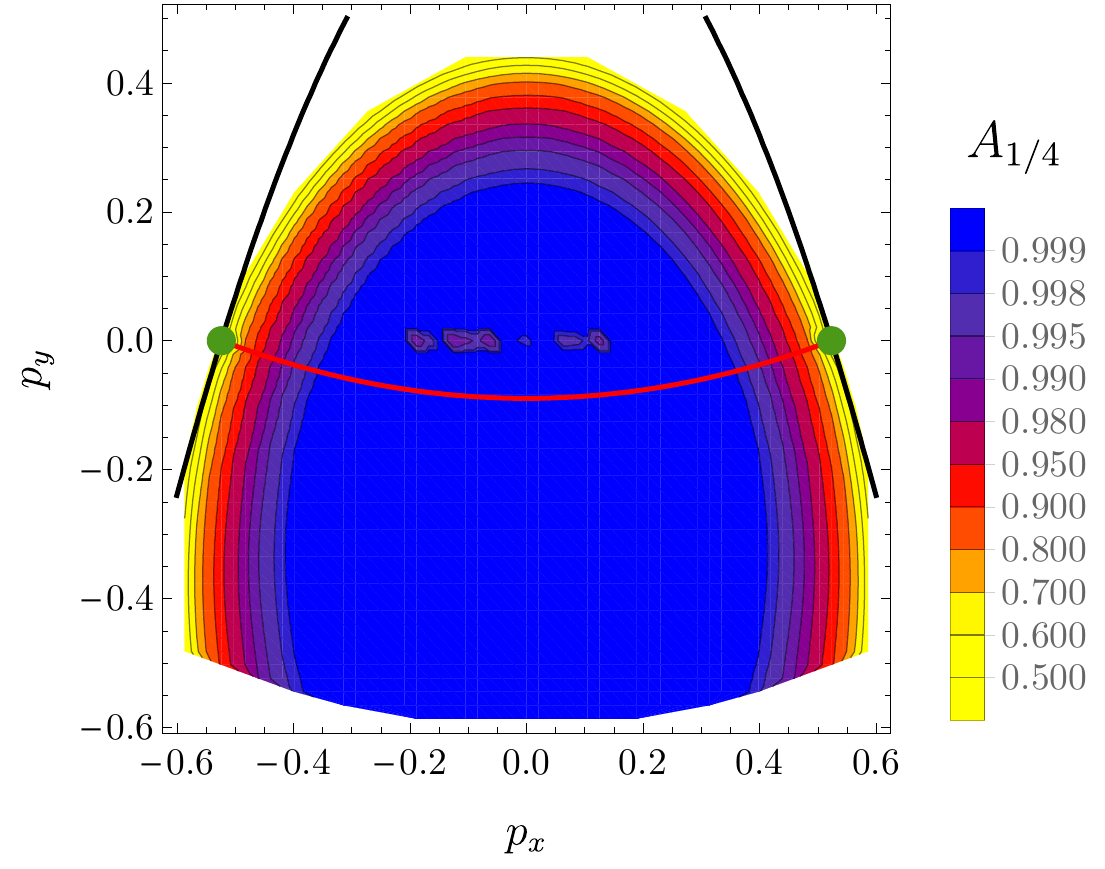}\llap{
  \parbox[b]{15.5cm}{(a)\\\rule{0ex}{6cm}}}
\hspace{0.01\textwidth}
\includegraphics[height=0.35\textwidth]{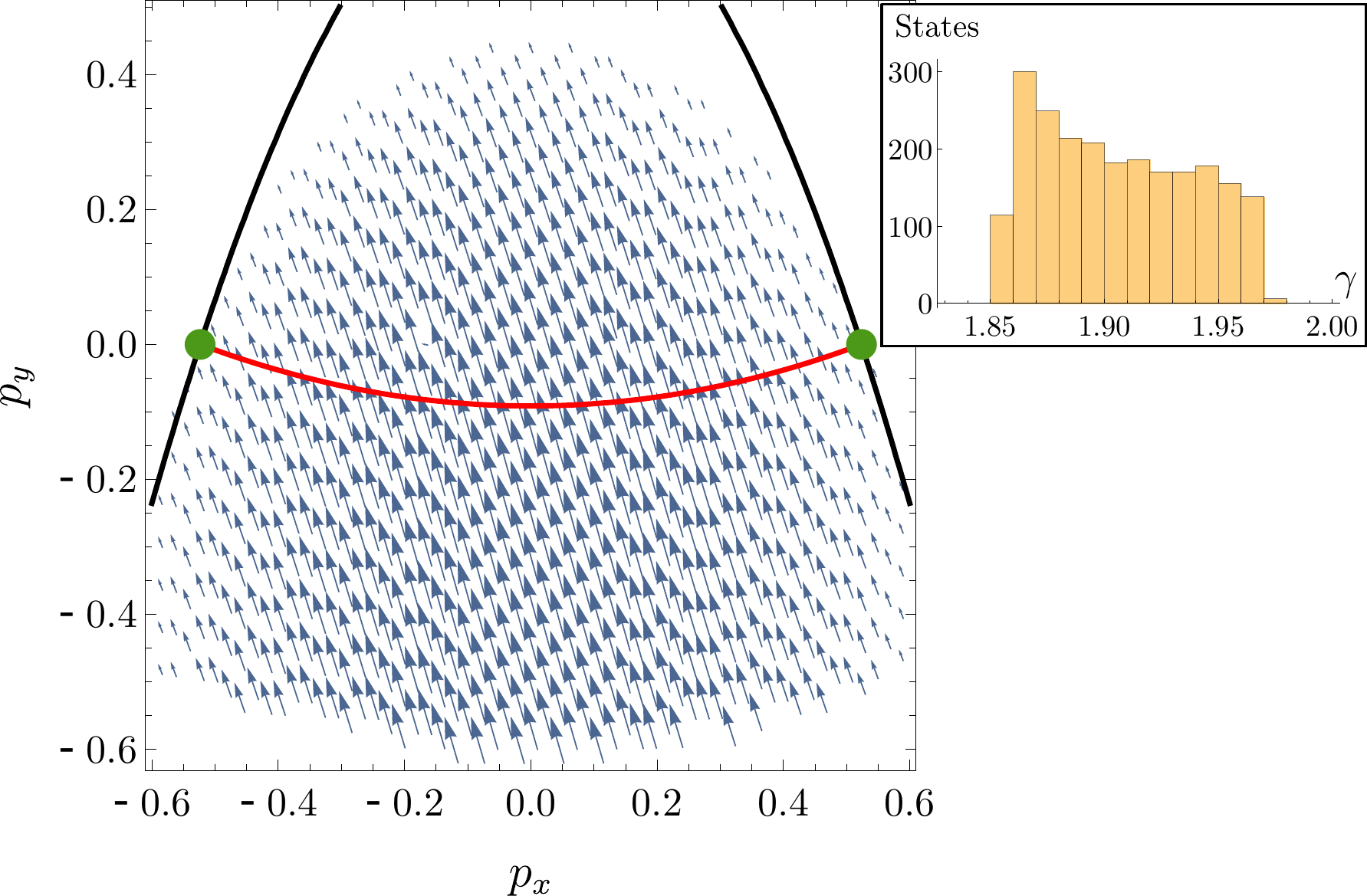}\llap{
  \parbox[b]{19cm}{(b)\\\rule{0ex}{6cm}}}
\\
\includegraphics[height=0.35\textwidth]{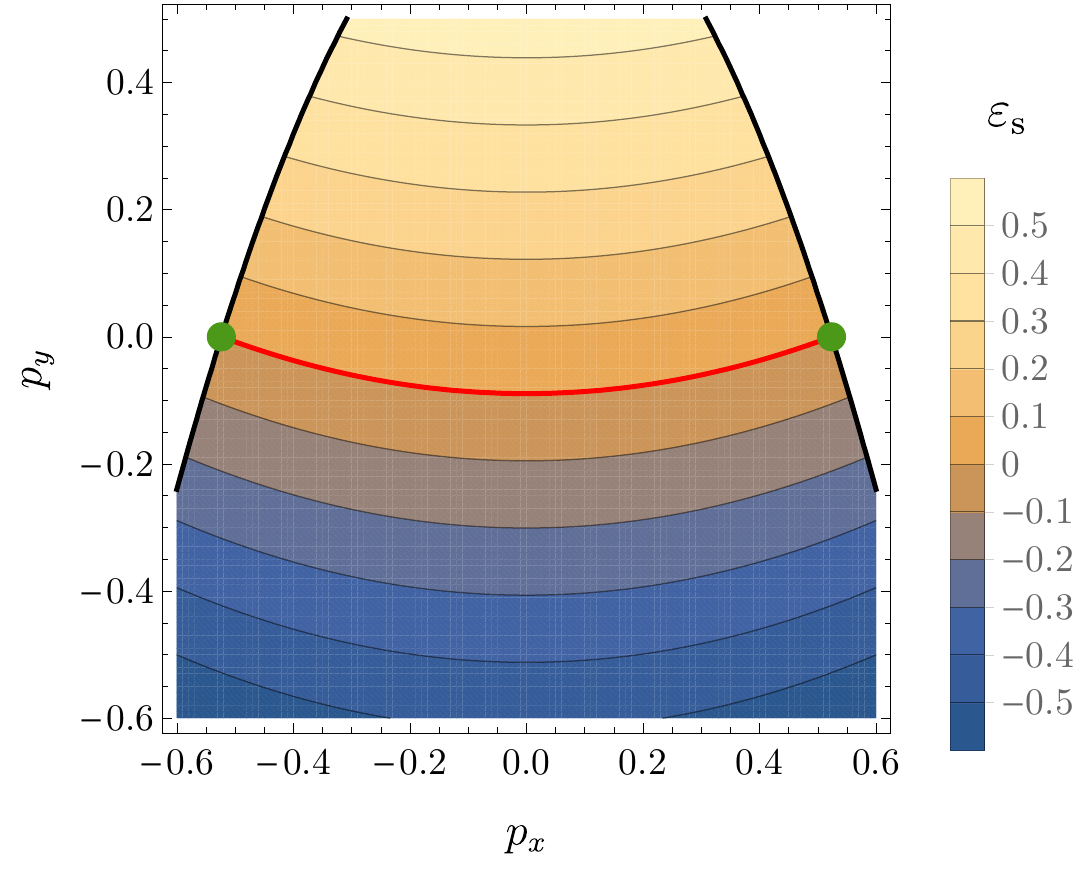}\llap{
  \parbox[b]{15.5cm}{(c)\\\rule{0ex}{6cm}}}
\hspace{0.01\textwidth}
\includegraphics[height=0.35\textwidth]{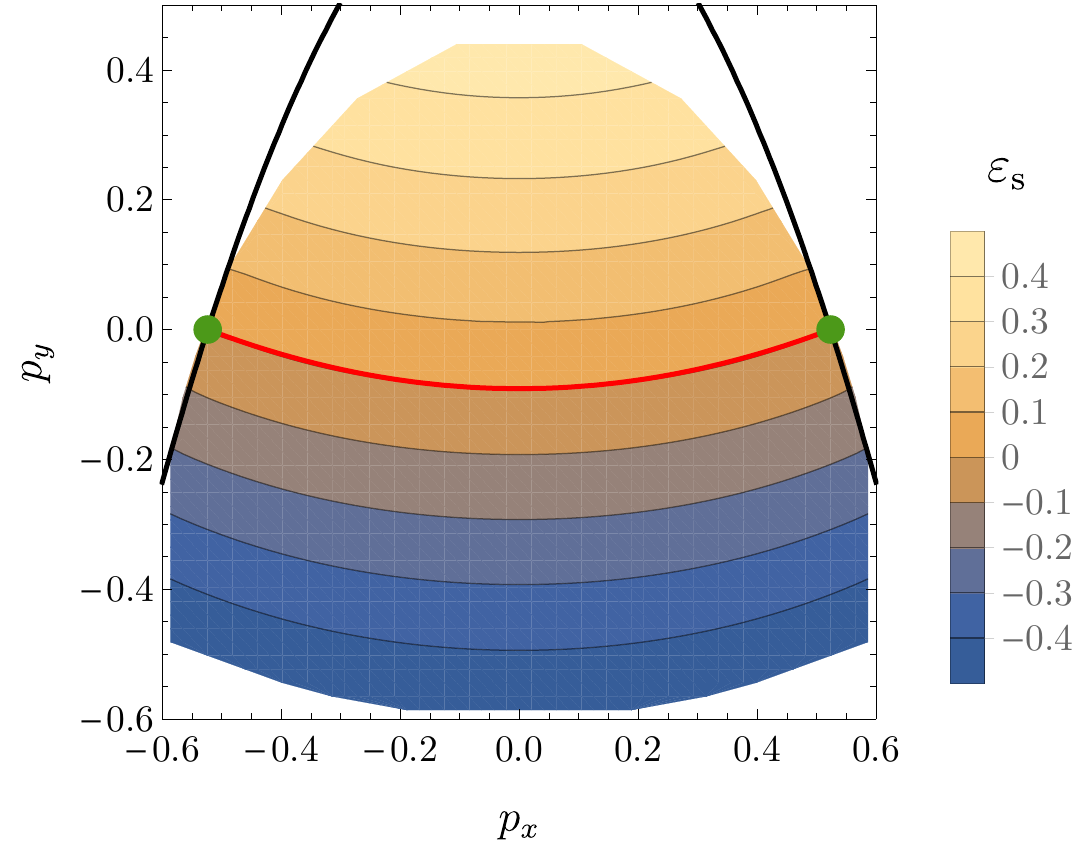}\llap{
  \parbox[b]{15.5cm}{(d)\\\rule{0ex}{6cm}}}
\end{flushleft}
\caption{Properties of the surface states for a surface Zeeman term \eqref{surfzeeman} with amplitude $B=0.2 v$ for $p_0=\pi/6$. The numerical data are obtained for a system of size $300 \times300 \times  90$  from the lattice model in Eq. \eqref{latHam} with periodic boundary conditions along $\hat{x}$ and $\hat{y}$. In all panels the black lines delimit the domain of the surface states based on the constraint $\tilde{p}_z>0$ (see Eq. \eqref{g}) for $\gamma=1.906$, the red lines depict the shape of the Fermi arc in Eq. \eqref{fa1} for the same value of $\gamma$, and the green dots correspond to the projection of the two Weyl points on the surface Brillouin zone. (a) Squared amplitude $A_{1/4}$ of the wavefunctions of the surface states evaluated for $z\le 20$; only states with $A_{1/4}>0.4$ and $|\varepsilon_{\rm s}| < 0.5v$ have been considered.  (b) Surface polarization of a subset of the surface states with $A_{1/4}>0.4$. $\gamma$ varies weakly in the surface Brillouin zone with values typically in the range $(1.85,1.97)$ and average $\gamma=1.906$ (see inset). (c) Energy of the surface states evaluated from Eq. \eqref{s2}. (d) Energy of the surface states calculated numerically; the analytical description in (c) matches well the numerical results in (d) for energies close to 0.} \label{fig:archi}
\end{figure*}

The results of the previous sections rely on the value of the surface polarization $\gamma$ defined in Eq. \eqref{bc}. Therefore, as a first step, we measure the value $\gamma (p_0)$ for several values of $p_0 \in [0, \frac{\pi}{2}]$. To this purpose we consider open boundary conditions along the $\hat{z}$ direction and periodic boundaries along $\hat{x}$ and $\hat{y}$ directions, in order to have only surfaces orthogonal to $\hat{z}$. In particular, we analyze systems with dimensions $L$ and $W\equiv W_x=W_z$ up to 150 and we estimate the parameter $\gamma$ by evaluating the surface polarization:
\beq
\gamma = \arctan \frac{\langle c^\dag_{\bf r_0} \sigma_y c_{\bf r_0}\rangle}{\langle c^\dag_{\bf r_0} \sigma_x  c_{\bf r_0} \rangle} \, ;
\label{formulagamma}
\eeq
here the expectation value is taken over the single-particle eigenstate of \eqref{latHam} corresponding to the lowest-energy eigenstate with positive energy. We verified that this state corresponds to a linear superposition of states localized on the two surfaces at $z \gtrsim 1$ and $z \lesssim W_z$ and it belongs to the (hybridized) Fermi arcs for the system with this geometry.
In particular, we considered the site at ${\bf r_0}=\left(\frac{W}{2}, \frac{L}{2}, 1\right)$ on the surface at $z=1$. We verified that the polarization $\gamma$ does not depend on the $x$ and $y$ coordinates when considering periodic boundary conditions in these directions and we checked that its dependence on $y$ is very weak also for open boundary conditions.

The relation \eqref{formulagamma} is easily obtained from the explicit form of the states localized on the surface at $z=0$ in the continuum model of the previous sections (see  Eqs. \eqref{bc} and \eqref{A1}). In the entire range $p_0 \in [0, \frac{\pi}{2}]$, we notably find only small deviations around the value $\gamma=\pi/2$, of the order of $10^{-2}$. This suggests that, in the thermodynamic limit, the lattice model \eqref{latHam} is indeed defined by the boundary polarization $\gamma=\pi/2$. Therefore, to be able of varying the parameter $\gamma$ and study its effect on the surface states, we introduce the following additional surface terms to the Hamiltonian:
\begin{equation} \label{surfzeeman}
H_{\rm s} = B \sum_{x,y}\left[ c^\dag_{x,y,1} \sigma_y c_{x,y,1} - c^\dag_{x,y,W_z}  \sigma_y c_{x,y,W_z}\right].
\end{equation}
These surface interactions correspond to opposite Zeeman-like terms for the pseudospin of the system aligned along the $\hat{y}$ direction and localized on the two surfaces at $z=1$ and $z=W_z$. We adopted only Zeeman fields in the $\hat{y}$ directions because we verified that, for analogous values of the coupling constants, the Zeeman terms along $\hat{x}$ cause a much weaker effect on the surface eigenstates. 

\textcolor{black}{The engineering of such a surface term in physical systems strongly depends on the material properties. When the pseudospin is related to the occupation of specific orbitals within a unit cell, then, depending on the lattice termination, suitable weak voltage gates on the surface may favor or disfavor these occupations, thus giving rise to such effective Zeeman-like terms for the pseudospin on the surface. Effective surface Hamiltonians of the kind \eqref{surfzeeman} can be obtained by the detailed analysis of the electrostatic properties of the interfaces between Weyl semimetals and other materials (or gates) which has been addressed in several works (see, for instance, Ref. \onlinecite{goerbig17}). Moreover, in the case of Weyl semimetals breaking time-reversal symmetry, it is also possible that the pseudospin is associated to the physical spin of the electrons; in such a situation, ferromagnets in contact with the surface may constitute useful tools for the engineering of such a surface Hamiltonian.}

\textcolor{black}{Our approach is dictated also by a more mathematical reason:} we observe that any lattice Hamiltonian with a finite number of degrees of freedom, such as $H_{\rm lat}$, is self-adjoint without the necessity of specifying any extension through boundary conditions. \textcolor{black}{Thus,} we introduce the surface Hamiltonian $H_{\rm s}$ to compensate for the lack of this freedom in parameterizing the boundary conditions and we verify in the following that, indeed, the surface terms \eqref{surfzeeman} modify the physical system in such a way that the analytical low-energy description presented in the previous sections provides an accurate approximation for the behavior of the system. In particular, we verify that, also for $B\neq 0$, the surface polarization is linked to the shape of the Fermi arcs and the density of surface states as a function of the energy, thus to $\sigma_H$.   

The properties of the surface states for $B=0.2v$ are illustrated in Fig. \ref{fig:archi}. These results correspond to periodic boundary conditions along $\hat{x}$ and $\hat{y}$ and are obtained by diagonalizing the Hamiltonian in the subspaces defined by the conserved momenta $p_x$ and $p_y$. To select the surface states, we considered a threshold $A_{1/4}$ corresponding to the squared amplitude of the wavefunctions in the interval $z \in [0,W_z/4]$. The results in Fig. \ref{fig:archi} correspond to all the states fulfilling $A_{1/4}>0.4$, thus sufficiently localized close to $z=1$. This threshold is however arbitrary and, in the comparison with the analytical low-energy model, the states we selected from the numerical simulation are only a subset of the corresponding surface states fulfilling the constraint $\tilde{p}_z>0$. Hence we expect some deviation of the numerical results from the analytical predictions for values of $\tilde{p}_z$ close to zero, thus close to the Weyl point projections and, more in general, close to the edges of the region where the analytical model predicts the existence of surface states, which are depicted as black lines on the surface Brillouin zone in all the panels of Fig. \ref{fig:archi}. Fig. \ref{fig:archi}(a) illustrate the probability $A_{1/4}$ for the eigenstates of the lattice model: in general, the amplitude $A_{1/4}$ decreases by approaching the predicted boundaries (black lines) and the Weyl point projections (green dots);  small irregularities can be observed for $p_y=0$ due to the hybridization of the surface states in opposite surfaces. For increasing values of $|p_y|$, thus of $|\varepsilon_s|$, the domain of the selected numerical surface states is smaller than the analytical prediction; this is mostly due to the differences for energies comparable with $v p_0$ between the low-energy analytical model and the lattice model. The discrepancies in the domains is also partially due to the constraint $A_{1/4}>0.4$ which implies an underestimation of the surface states domain in the numerical data.

Fig. \ref{fig:archi}(b) displays the polarization \eqref{formulagamma} of the surface eigenstates. The value of $\gamma$ weakly varies as a function of the conserved momenta, and for $B=0.2v$, its average is $\gamma \approx 1.906$. We adopted this value in the analytical determination of the surface state domain and of the shape of the Fermi arc. The Fermi arc derived by the analytical prediction \eqref{fa1} matches very well the numerical results and, in general, for energies close to zero, the agreement between analytical and numerical models is very good, as shown by the comparison of the panels (c) and (d).

\begin{figure}[t]
\includegraphics[width=\columnwidth]{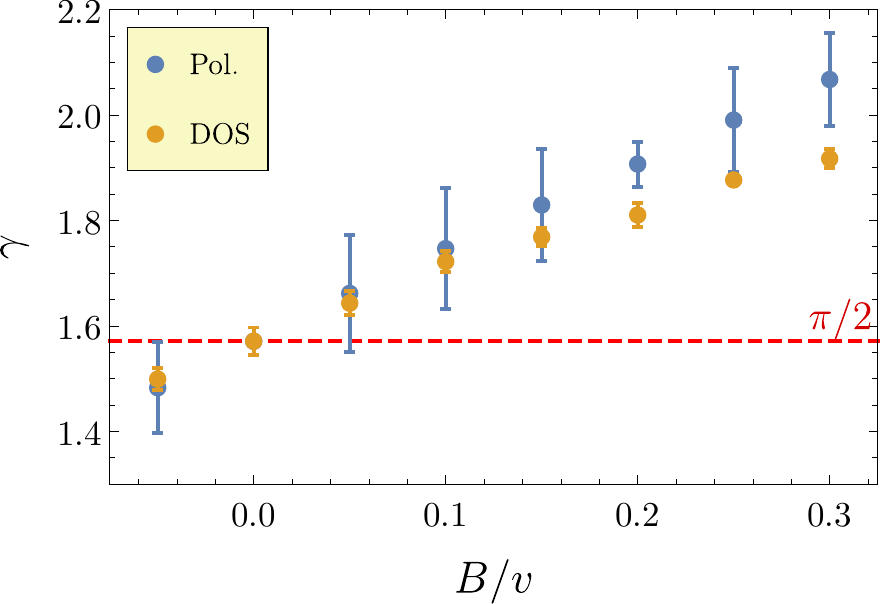}
\caption{Values of $\gamma$ determined by the surface polarization (blue points) and the surface density of states (yellow points) as a function of the boundary term \eqref{surfzeeman}. The error bars correspond to the standard deviation and to the standard error in the fitted parameter, respectively.}
\label{fig:err}
\end{figure}

Fig. \ref{fig:err} displays the values of the surface parameter $\gamma$, as a function of the boundary field $B$. We compare two different ways of obtaining the estimate of $\gamma$: (i) we consider the average of the expectation value of the polarization \eqref{formulagamma} for the surface states, selected based on the constraint $A_{1/4}>0.4$ for the energy interval $\varepsilon_{\rm s} \in \left[-0.5v,0.5v\right]$; (ii) we consider the value of $\gamma$ obtained from a fit of the density of surface states.

Concerning the approach (ii), the number of surface states for an energy interval $d\varepsilon_{\rm s}$ is derived from Eqs. \eqref{s2} and \eqref{g} and it results in:
 \begin{equation} \label{doss}
N(\varepsilon_{\rm s}) = \left(\frac{W_xL_y}{2\pi^2v\sin\gamma}\sqrt{p_0^2 + 2p_0\frac{\varepsilon_{\rm s}}{v}\cos \gamma}\right)d\varepsilon_{\rm s}\,.
\end{equation}
As mentioned in Section \ref{zeroT}, this quantity differs from the particle number on the surfaces, defined via Eq. \eqref{rhoT}.
  
\begin{figure}[t]
\includegraphics[width=\columnwidth]{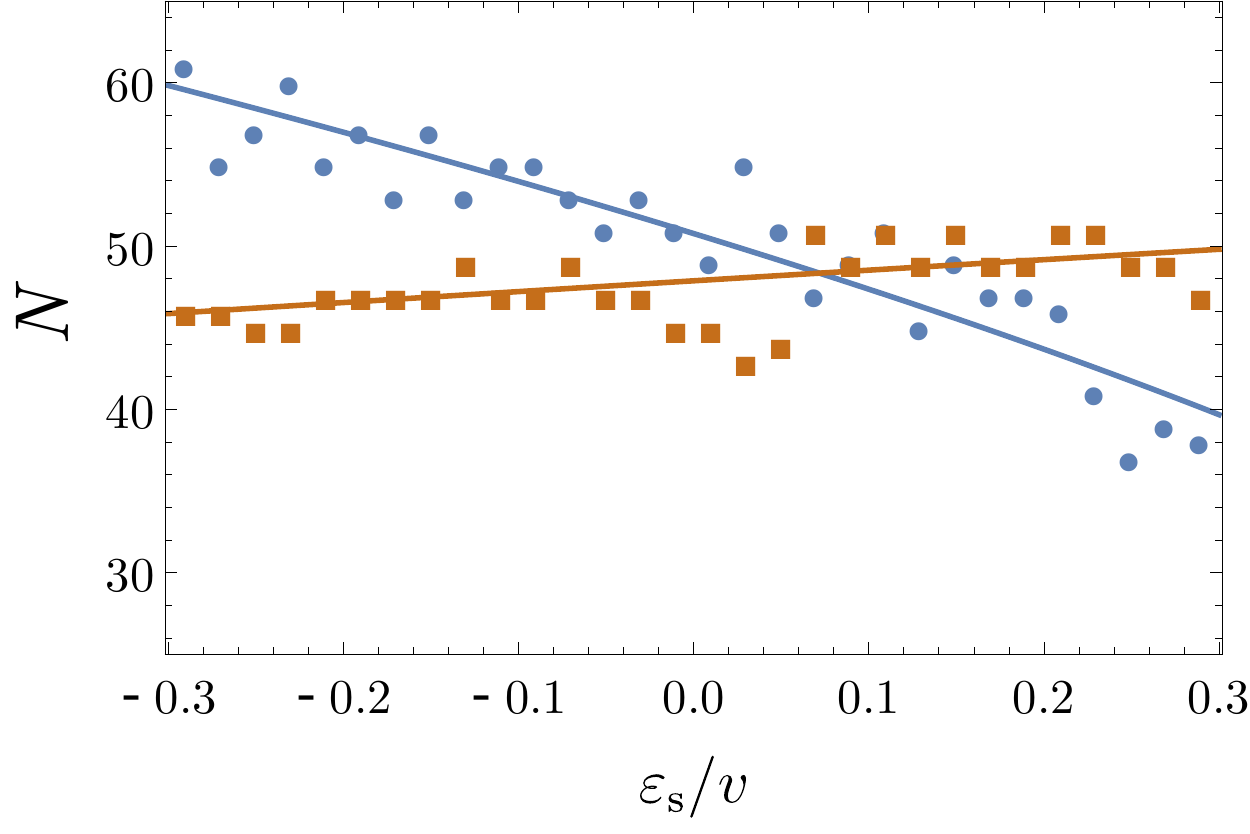}
\caption{Number of surface states $n(\varepsilon_{\rm s})$ (see Eq. \eqref{doss}) with $A_{1/4}>0.4$ for a lattice system of dimension $300\times 300 \times 90$ and surface Zeeman fields $B=0.3v$ (blue circles) and $B=-0.05v$ (brown squares). The numerical data correspond to energy intervals with $d\varepsilon_{\rm s}=0.02v$. The curves illustrate the result of fits with $\gamma$ as the only fitting parameter.}
\label{fig:fit}
\end{figure}

We use Eq. \eqref{doss} to perform a one-parameter fit of the numerical data concerning the number of states with $A_{1/4}>0.4$ of a system with dimension  $300\times 300 \times 90$ for the energy range $\varepsilon_{\rm s} \in \left[-0.3 v,0.3 v\right]$ with intervals $d\varepsilon_{\rm s}=0.02v$. Two examples of the fit result and density of surface states are shown in Fig. \ref{fig:fit} for $B=0.3v,-0.05v$. This method is potentially affected by a larger systematic error because of the lattice and finite size effects, that
determine considerable oscillations of the density for the surface states and cause a deviation from the analytical model for $\epsilon_s$ approaching $vp_0$.

The numerical results above are obtained for an isolated system with periodic boundary conditions in the $\hat{x}$ and $\hat{y}$ direction. In the following we focus on the transport properties of the Weyls semimetal in contact with two external leads. 

In particular, to probe the surface conductivity, we include two semi-infinite semimetallic leads in the system, and we connect them to the Weyl scattering region on the facets at $y=1$ and $y=L_y$, thus reproducing the geometry of the previous section. These leads are described by the same Hamiltonian \ref{latHam} and they are approximately characterized by the same chemical potential of the scattering region.

This configuration is required to probe directly the conductivity of the scatterer, avoiding non-universal effects from the leads and the interfaces between leads and scattering region: such aspect is especially important for the dynamics of surface states. In more detail, to probe the low-energy physics around the Weyl nodes, we set the chemical potential of the leads $\mu_{\mathrm{lead}} = 0$, and in the scattering region $\mu_{\mathrm{bulk}}$  at a slightly larger value (typically $\mu_{\mathrm{bulk}}/v \approx 10^{-3} - 10^{-2}$). To simplify our numerical calculations, we maintained periodic boundary conditions in the $\hat{x}$ direction, thus diagonalizing the system in different Hilbert space sectors labeled by $p_x$, and we considered system sizes with $W_x=W_z \equiv W$.

\begin{figure}[t]
\includegraphics[width=\columnwidth]{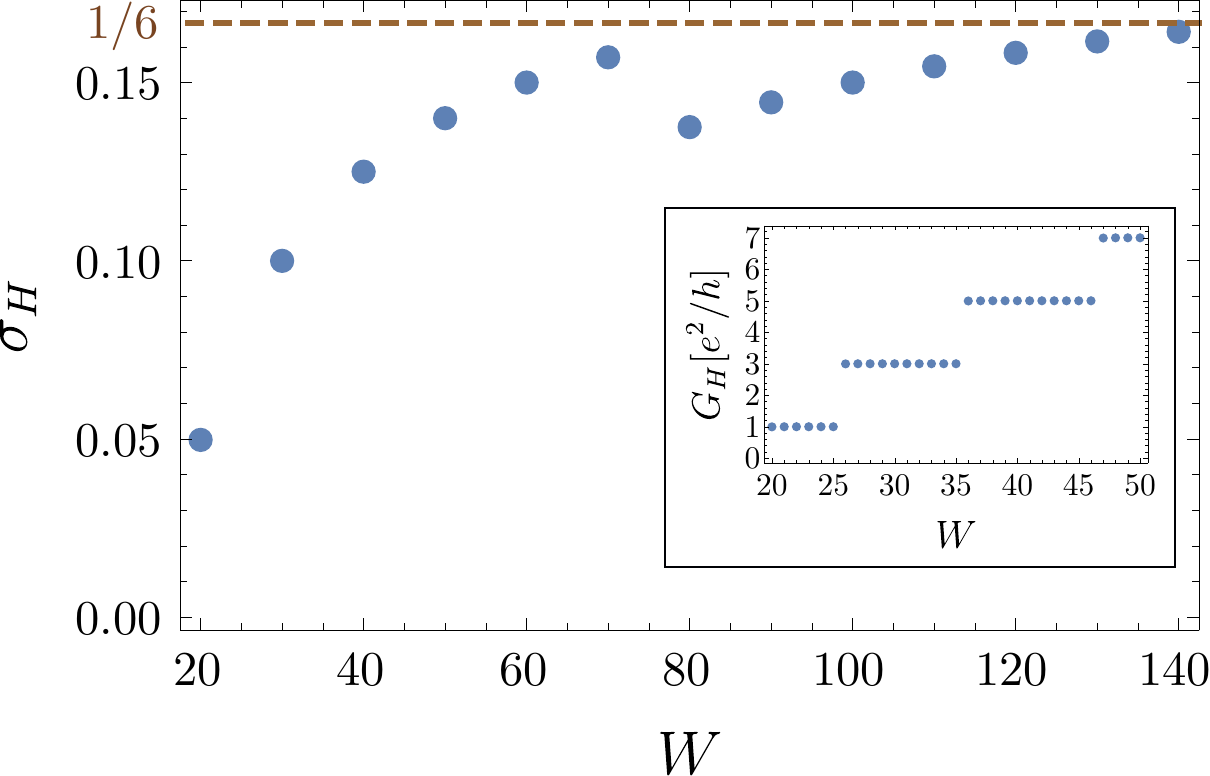}
\caption{Surface conductivity (in units of $e^2/h$) of a system with $L = 30$, 
   and $p_0=\pi/6$, as a function of $W$. 
The numerical calculation is performed at energy $\mu_{\mathrm{bulk}}=10^{-3}v$ and $\mu_{\rm lead}=0$, with periodic boundary conditions along $\hat{z}$ and Weyl semimetallic leads. For large system sizes, the Hall conductance approaches the expected universal value $(1/6) e^2/h$. Inset: quantized Hall conductance $G_H$.}
\label{sigmahnum}
\end{figure}

The resulting conductance of the system corresponds, in the
thermodynamical limit, to the anomalous Hall conductance $G_H = W \,  \sigma_H$.
This is a direct consequence of our choice of the leads, with the same
Hamiltonian and vanishing chemical potential as the scatterer:
the bulk density of states vanishes also in the leads, and the only
contribution to the total conductance is given by the surface states.
Under these conditions, even for small values of $L_y$ and $W$, we observe a clear quantization of $G_H$ in units of $e^2/h$ (see the inset of Fig. \ref{sigmahnum}), analogously with the behavior of Weyl semimetal nanowires \cite{bardarson19}. Such quantization corresponds to the number of states in the Fermi arc at energy equal to $\mu \approx 0$; for the data of Fig. \ref{sigmahnum}, $G_H$ assumes odd values of the conductance quantum.  For the systems considered in the present Section, the total number of states  in the Fermi arc for $\mu \approx 0$ and $0< \gamma < \pi$ results: $N^{(tot)}_s  = \Big[2 p_0/ (\frac{2 \pi}{W_x})\Big] = \Big[p_0 W_x/\pi \Big]$, $\big[ \big]$ denoting the integer part. This value leads to the universal anomalous Hall conductance in Eq. \eqref{twocond}.
 
For sufficiently large $W$, thus with negligible hybridization of the surface states on the opposite surfaces, we find that the Hall conductivity $\sigma_H = {G_H}/{W}$ tends to $1/6 = {p_0}/{\pi}$, the universal value predicted in Eq. \eqref{hall} at $\mu = 0$ (see Fig. \ref{sigmahnum}). We observe, however, that $\sigma_H$ presents some discontinuities as a function of the width of the system, due to finite size effects. For the largest system size we probed, $W=140$, we obtained $\sigma_H(W  = 140) \approx 0.164 e^2/h$.

We conclude by analyzing the transport properties of the bulk. To this purpose we consider systems with periodic boundary conditions in the $\hat{x}$ and $\hat{z}$ directions. By maintaining vanishing chemical potentials in both the leads and the scatterer, we calculated the bulk conductance $G_{\rm b}$ for $p_0=\pi/6$, in both the fine-tuned and standard regimes (the fine-tuned regime is given by $W=12n$ with $n \in \mathbb{N}$, such that $p_0 W$ is a multiple of $2\pi$). For example, we consider $W = 24$ and $W  = 25$. 
In the fine-tuned regime, already for $L_y = 10$ and $W=24$, we measure $G_{\rm b} = 2 {e^2}/{h}$, up to an error of $10^{-13}$. This verifies the existence of the two expected zero-energy non-evanescent bulk states, which correspond to the Weyl band-touching points and match the continuum model description. However, differently from the previous section and the calculation in Eqs. \eqref{transmission} and \eqref{transsum}, our choice of vanishing chemical potential in the leads implies a vanishing of their density of states, such that there is no other contribution to the transport in this regime apart from these two zero-energy bulk states.

A comparison with Eqs. \eqref{transmission} and \eqref{transsum} can be performed, instead, by adopting metallic leads, which may be simply modeled by a cubic lattice Hamiltonian of fermions with spin-independent nearest-neighbor hopping terms; in this case a high density of states occurs in the leads, similarly to the Landauer-Buttiker approach of  Eq. \eqref{transsum} (where the limit $\mu_{\mathrm{lead}} \to \infty$ is adopted).

\begin{figure}[t]
\includegraphics[width=\columnwidth]{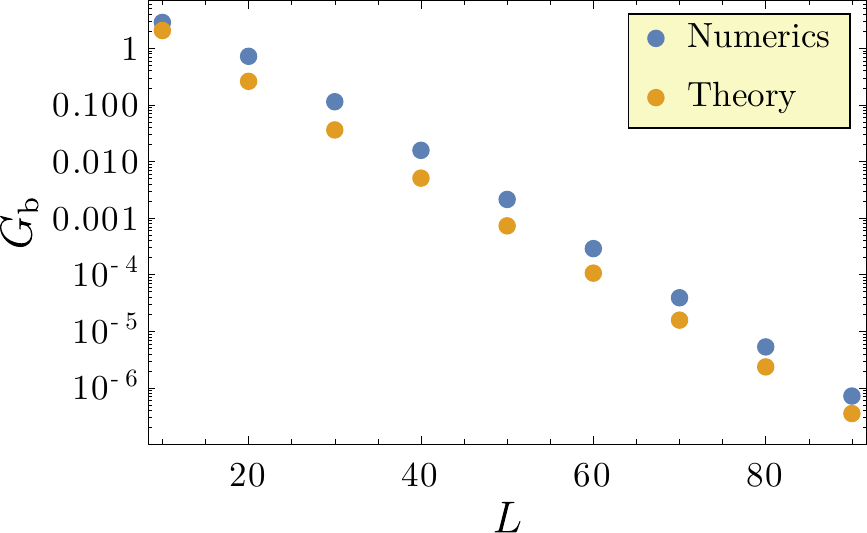}
\caption{Bulk conductance in Eq. \eqref{conddens} for $W = 30$, $\mu_{\mathrm{bulk}}= 10^{-3}$, and $p_0=\pi/6$, as a function of $L$  and in logarithmic scale. The yellow dots are the results obtained from Eq. \eqref{transsum}, whereas the blue dots denote the numerical results for a scatterer described by the lattice Hamiltonian \eqref{latHam} with metallic leads and periodic boundary conditions along $\hat{x}$ and $\hat{z}$. For large $L$, the analytical results typically underestimate the numerical data by a factor $\sim 3$.}

\label{comparison}
\end{figure}

The analytical estimate of the bulk conductance in Eq. \eqref{transsum} is compared with the numerical result with metallic leads in Fig. \ref{comparison} for $W = 30$, $\mu= 10^{-3}$, and $p_0=\pi/6$, as a function of $L$. The yellow dots represent the results from Eq. \eqref{transsum}, while the blue dots denote the numerical data. The analytical results underestimate the numerical conductance typically by a factor $\sim 3$, thus it provides only an estimate of the order of magnitude of the bulk conductance. This discrepancy is due to the different kinds of leads considered, Weyl semimetallic leads for the analytical estimate and metallic leads for the numerical results, and to additional interface effects that can stem from different boundary conditions at the interface between the wavefunctions of the leads and the scatterer than the ones considered in Appendix \ref{app:conductance}.

\section{The effects of temperature} \label{secT}

\subsection{Thermal noise}

The field theoretical approach described in Sec. \ref{zeroT} can be used to obtain an estimate of the current and conductivity behavior also as a function of temperature. We saw that, for $\gamma=\pi/2$, the surface current $\left\langle j_y\right\rangle$ does not depend on $T$ and $\mu$, whereas the average value of the currents along $x$ and $z$ always vanishes, as dictated by the symmetries of the system. 
In the ballistic regime, the chiral surface modes can be considered in thermal equilibrium with the leads they originate from; therefore, for the geometry considered in the previous section, we can introduce surface chemical potential $\mu_{\rm s}$ and inverse temperature $\beta_{\rm s}$ equal to the parameters of the lead at $y<0$ for $\sin \gamma >0$ (again $0<\gamma<\pi$ ). In the case $\sin \gamma <0$, which is in general not well-defined in our continuum model, the surface current changes direction and  $\mu_{\rm s}$ and $\beta_{\rm s}$ would instead be derived from the lead at $y>L$.

From the definition of the current operator, we can derive the spectral density of the surface current noise at frequency $\nu$:
\begin{multline} \label{snoise1}
S_{yy}\left(\nu,\bf{r_1},\bf{r_2}\right)=\\ \int_{-\infty}^\infty d t\,  \e^{-i \nu t} 
\left[\langle j_{y} (t,{\bf r_1})  j_{y} (0,{\bf r_2})\rangle -
\langle j_{y} (t,{\bf r_1})\rangle \langle  j_y (0,{\bf r_2})\rangle\right]\,.
\end{multline}
For $\gamma=\pi/2$, the calculation can be explicitly done. In the limit $\nu\to 0$, for $x_1=x_2$ and $z\equiv z_1=z_2$, we obtain:
\begin{multline}
S_{yy} =   \frac{p_0} {4 \pi^3 z^3 \beta_s \left (1+\e^{-\beta_{\rm s}\mu_{\rm s}}\right )} \times \\ 
\left [(1+2p_0 z)\, D_+\left(\sqrt {p_0 z})\right) - 
\sqrt {p_0 z}\, \right]^2\, ,  
\label{snoise2}
\end{multline}
which does not depend on $y_1$ and $y_2$ and decays asymptotically as $z^{-4}$.
The shot noise vanishes,  
\begin{equation}
\lim_{\beta \to \infty}S_{yy} =  0\, ,  
\label{n3}
\end{equation}
consistently with the chiral nature of the surface states. The pure thermal limit gives instead:  
\begin{equation}
\lim_{\mu_{\rm s} \to 0}S_{yy} =   \frac{p_0} {8 \pi^3 z^3 \beta_s } \left [(1+2p_0 z)\, D_+\left(\sqrt {p_0 z})\right) - 
\sqrt {p_0 z}\, \right]^2 ,  
\label{n4}
\end{equation}
which respects the Johnson-Nyquist law. 

Eq. \eqref{snoise2} is the two-point correlation function of the surface current density only. A complete calculation of the noise spectrum, however, must take into account also the bulk-bulk and bulk-surface correlations at different points. 

\subsection{Bulk currents}

The expectation value \eqref{jyexp} accounts exclusively for the surface current; the total current density is given by 
\begin{align}
J_x(t, \bm r) &= \frac{iv }{2p_0} :\left [ (\der_x \Psi^\dagger )\sigma_x \Psi - \Psi^\dagger \sigma_x (\der_x \Psi )\right ] : (t, \bm r)\, ,  \\
J_y (t,\bm r) &= v  :\Psi^\dagger \sigma_y \Psi : (t,\bm r)\, ,  \label{Jtoty}\\
J_z (t,\bm r) &= v  :\Psi^\dagger \sigma_z \Psi : (t,\bm r)\, ,
\end{align}
and includes both surface and bulk contributions. Employing the two-point functions 
(\ref{fermi1}-\ref{fermi4}) and the explicit form (\ref{A1},\ref{A2}) of the surface and bulk eigenfunctions one obtains the 
following mean values: 
\be
\left\langle J_x(t,{\bm r})\right\rangle = \left\langle J_z(t,{\bm r})\right\rangle = 0\, , 
\label{currenttot1}
\ee
\begin{multline} 
\label{currenttot2}
\left\langle J_y(t,{\bm r})\right\rangle = v\sin(\gamma) \int \frac{d^2 p}{(2\pi)^2} \Theta[\pt_z ({p})]2 \pt_z ({p}) \e^{-2\pt_z ({p}) z} \times \\
\left\{\frac{1}{1+\e^{\beta_{\rm s} \left[\varepsilon_{\rm s}({p})-\mu_{\rm s}\right]}} - \frac{1}{1+\e^{\beta_{\rm b} \left[\varepsilon_{\rm s}({p})-\mu_{\rm b}\right]}}\right\}\, . 
\end{multline} 
The surface state contribution gives the first term in the curly brackets. The original contribution of the bulk states (\ref{A2}) involves an integration over $d^3p$. 
Using the Cauchy integral formula and the pole structure of the scattering matrices (\ref{A4},\ref{A5}), one can perform the $p_z$ integral, which leads to 
the second term. In this operation one uses the fact that restricting $p_z$ in $\varepsilon_{\rm b}({\bf p})$ to the poles of (\ref{A4},\ref{A5}) 
one gets precisely $\varepsilon_{\rm s}({p})$. For this reason the surface energy $\varepsilon_{\rm s}$ enters both the bulk and surface contributions.

In general, the assumption of having different temperatures and chemical potentials for the surface and bulk states is justified in the ballistic regime: our model neglects scattering terms and interactions between bulk and surface states. Such terms naturally appear in a physical system due to disorder \cite{gorbar16} or electron-phonon coupling \cite{resta18}  and determine a relaxation time $\tau$ beyond which bulk and surface states equilibrate. However, if $\tau \ll L/(v\sin \gamma)$, we may assume that a quasiparticle does not equilibrate during the transport between the two leads in the geometry discussed in the previous section. In this situation, as a first approximation, it is legitimate to choose different bulk and surface Fermi distributions. 

If instead we consider the system with $\beta_{\rm s} = \beta_{\rm b}$ and $\mu_{\rm s}=\mu_{\rm b}$, 
from (\ref{currenttot2}) one immediately infers that the mean value of total current $\left\langle {\bm J}(t,{\bm r})\right\rangle$ vanishes everywhere. 
In this case we are dealing in fact with the thermodynamic limit of a system at equilibrium. 
This result may not hold in finite systems: by including two parallel surfaces in our model with a finite separation, thus two 
different scattering matrices, the expectation value of the local current density at equilibrium may in general be 
different from zero and depend on position, consistently with previous numerical results displaying persistent currents 
in small Weyl semimetals \cite{lopez2018,zhang2018}.

Also concerning the spectral density of the noise, at equilibrium, the correlations between bulk operators provide a contribution identical to the surface states, thus doubling the result \eqref{n4}. Additionally, one should consider also correlations between surface and bulk currents whose computation go beyond the scope of this work.

\subsection{Thermal Hall conductivity}

Heat currents are typically more difficult to measure than electric currents. In experimental systems the heat transport is determined not only by the electrons, but also by the phonons propagating in the material. When we restrict our attention to the heat transport along the Weyl semimetal surfaces, the electronic contribution to the heat current typically scales with $T^2$, due to the chiral dispersion of the Fermi arcs, whereas the phonons are free of propagating in any direction, leading to a typical Stefan-Boltzmann behavior proportional to $T^4$ (or $T^5$ for more refined models \cite{wellstood1994}). Therefore, analogously to the heat transport in quantum Hall setups, the electronic contribution dominates for low temperatures; we conclude that the calculation of the electronic heat conductivity of our model provides an approximate description of the heat transport for low temperatures. 

Similarly to the charge transport \cite{umansky2017}, Weyl semimetals display an anomalous Hall effect also for the heat transport\cite{goswami2013,Lundgren2014}. Former analysis based on the bulk properties of Weyl semimetals suggests the existence of a universal value of the thermal Hall conductivity, $\kappa_H=p_0 {\pi k_B^2 T_{\rm s}}/{3h}$ at vanishing chemical potential, such a value fulfills the Wiedemann-Franz law and is consistent with the energy transport of the chiral states on the surface \cite{kane1997,cappelli2002}. In the following we analyze the energy transport on the surface of our model and we show that, in general, the boundary conditions yield non-universal corrections of the thermal transport, determining, in turn, a violation of the Wiedemann-Franz law.

Based on the total field $\Psi$ and the surface field $\Psi_{\rm s}$, the total and surface energy currents of our system, in the $\hat{y}$ direction orthogonal to the Weyl point separation, are defined by:
\begin{align}
&\mathfrak{J}_y(t,{\bm r})= \frac{iv}{2}:\left[\Psi^\dag \sigma_y \partial_t \Psi - \partial_t \Psi^\dag \sigma_y \Psi \right]: (t,{\bm r})\,,\label{totheatcurrent}\\
&\mathfrak{j}_y(t,{\bm r})=\frac{iv}{2}:\left[\Psi_{\rm s}^\dag \sigma_y \partial_t \Psi_{\rm s} - 
\partial_t \Psi_{\rm s}^\dag \sigma_y \Psi_{\rm s} \right] :(t,{\bm r})\,. \label{sheatcurr}
\end{align}
The electric and energy currents (\ref{jyop},\ref{Jtoty},\ref{totheatcurrent},\ref{sheatcurr}) generate\cite{Callen} the bulk and surface heat currents, 
\begin{align}
&Q_y (t,{\bf r}) = \mathfrak{J}_y(t,{\bf r}) - \mu_{\rm s}\, j_y(t,{\bf r}) - \mu_b \, \left[J_y(t,{\bf r}) - j_y(t,{\bf r})\right]\,,\\
&q_y(t,{\bf r}) =\mathfrak{j}_y(t,{\bf r}) - \mu_{\rm s}\, j_y(t,{\bf r})\, .  
\label{heatcurrents}
\end{align}
Analogously to the case of the electric transport, one can extract the thermal Hall conductance 
$\kappa_H$ from the surface heat current \eqref{heatcurrents} as a function of the surface temperature $T_{\rm s}$. We obtain:
\begin{multline} \label{thermalhall}
{\kappa_H}(\mu,T_{\rm s}) = \partial_{T_{\rm s}}\int d{p_x} \int_0^\infty dz \left\langle \hat{q}_y(p_x,z,\mu)\right\rangle_{T_{\rm s}}=\\
= \frac{v\sin\gamma}{k_BT_{\rm s}^2} \int \frac{d^2p}{(2\pi)^2} \frac{\Theta(\tilde{p}_z)\left[\varepsilon_{\rm s}\left(p\right) -
\mu_{\rm s}\right]^2}{4\cosh^2\left[\frac{\varepsilon_{\rm s}\left(p\right) -\mu_{\rm s}}{2k_BT_{\rm s}}\right]}\,.
\end{multline}

In the case $\gamma=\pi/2$ of a straight Fermi arc, the Fermi arc dispersion depends only on $p_y$ and the previous expression simplifies. In particular, in the limit of vanishing chemical potential, we recover the predicted universal result \cite{goswami2013}:
\begin{equation} \label{universalthermal}
\kappa_H\left(T_{\rm s},\mu=0,\gamma=\pi/2\right) = \frac{\pi k_B^2 T_{\rm s}}{3h}p_0\,.
\end{equation} 
This result is consistent with the Wiedeman-Franz law, expected for non-interacting fermions in the absence of boundaries. For $\gamma=\pi/2$, indeed the density of surface states is constant, thus causing no corrections to the universal value \eqref{universalthermal} as a function of temperature or chemical potential. We verify the Wiedeman-Franz law by comparing the heat and electric currents. The two current densities are proportional to each other and, in particular, we get:
\bwt
\begin{align}
&\left\langle {J}_y (z)\right\rangle = \frac{\sqrt{p_0 z} \left[(1+2p_0 z)\, D_+(\sqrt {p_0 z}) - \sqrt {p_0 z}\, \right]} {4 \pi^2 z^2} \left[\mu_{\rm s}-\mu_{\rm b}\right]\,, \label{Jpi2} \\
&\left\langle Q_y (z)\right\rangle = \frac{\sqrt {p_0 z} \left [(1+2p_0 z)\, D_+(\sqrt {p_0 z}) - 
\sqrt {p_0 z}\, \right]} {24 \pi^2 z^2} \left [\left (\frac{\pi^2}{\beta^2_{\rm s}} +2\mu^2_{\rm s} \right ) -\left (\frac{\pi^2}{\beta^2_{\rm b}} +2\mu^2_{\rm b} \right )\right ], \label{heatcurr}
\end{align}  
\ewt
corresponding to a Lorentz number:
\begin{equation} \label{WF}
\mathcal{L}\equiv\frac{\kappa_H}{T_s\sigma_H}=\frac{\pi^2k_B^2}{3e^2}\,, \quad \text{for} \; \gamma=\frac{\pi}{2}\,,
\end{equation}
for both bulk and surface states.

\begin{figure}[tp]
\includegraphics[width=\columnwidth]{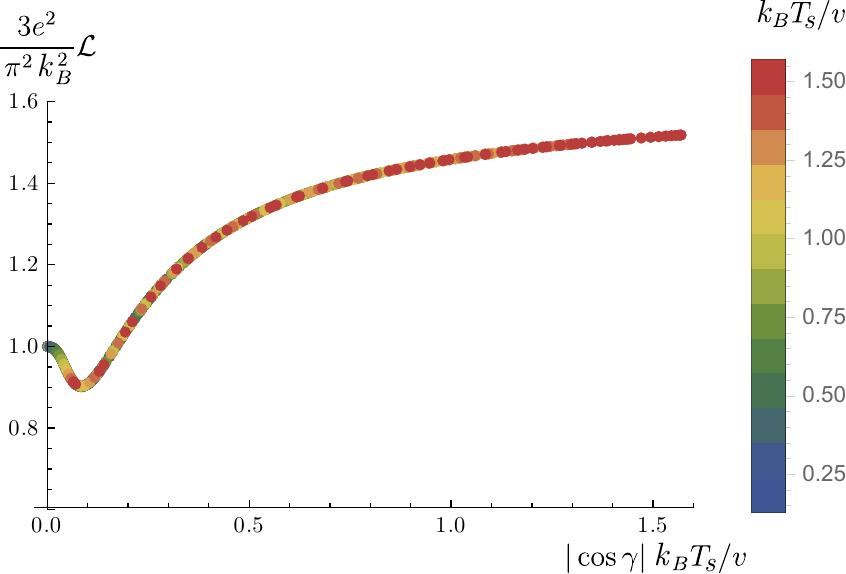}
\caption{Lorentz ratio $\mathcal{L}$, normalized by the parameter ${\pi^2k_B^2}/{3e^2}$, calculated for $\mu_s=0$ and $p_0=\pi/6$ by a numerical estimation of Eq. \eqref{thermalhall}. The curve has been obtained by estimating $\kappa_H$ for several values of the temperature (represented by the point colors) and the boundary angle $\gamma$ and plotting them as a function of the parameter $|\cos\gamma|k_BT_s/v$, consistently with Eqs. \eqref{alpha} and \eqref{Lorentzmaster} for $\mu_s=0$. The Lorentz ratio at $\mu_s=0$ depends non-trivially on the temperature for $\gamma\neq\pi/2$, and the Wiedemann-Franz law, $\mathcal{L} = {\pi^2k_B^2}/{3e^2}$ is recovered in the limit $\cos\gamma \to 0$.}
\label{fig:thermal}
\end{figure}

It is instructive to analyze the heat current \eqref{heatcurr}: its dependence from surface and bulk temperature is given by $T_{\rm s}^2-T_{\rm b}^2$. In a more realistic description we may consider a temperature that varies smoothly from the surface to the bulk, such that $T_{\rm s}^2-T_{\rm b}^2 \to - T \partial_z {T}/2$. Such substitution determines indeed a Hall heat current orthogonal to the gradient of the temperature.

\begin{figure}[tb]
\includegraphics[width=\columnwidth]{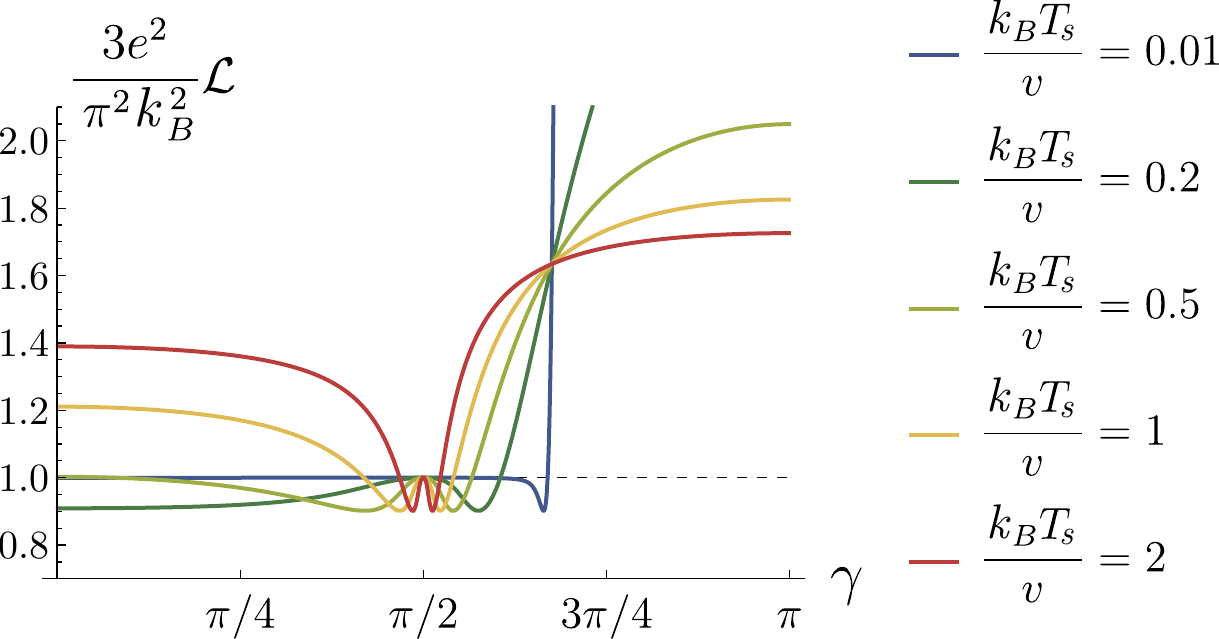}
\caption{Lorentz ratio $\mathcal{L}$ as a function of the boundary polarization $\gamma$. The Lorentz ratio is normalized by the parameter ${\pi^2k_B^2}/{3e^2}$ and is calculated from Eq. \eqref{Lorentzmaster} for values of $\alpha$ in Eq. \eqref{alpha} determined by $\mu_s/v=0.5$, $p_0=\pi/6$ and several values of the temperature. The blue curve at temperature $k_BT_s/v=0.01$ shows that, for large values of $\gamma$ (and positive chemical potential), the Wiedemann-Franz law is violated and the Lorentz ratio diverges. The crossing point of all the curves corresponds to $\alpha=0$ in Eq. \eqref{alpha}}
\label{fig:lorentzgamma}
\end{figure}

In the general case $\gamma \neq \pi/2$, the Lorentz ratio $\mathcal{L}$ of the Hall transport has a non-trivial dependence from the boundary polarization $\gamma$ (see Fig. \ref{fig:lorentzgamma}) and the behavior of the anomalous thermal Hall conductivity can be estimated numerically from the integral in Eq. \eqref{thermalhall}. At $\mu_s=0$ we numerically observe that the Lorentz ratio is a non-trivial function of the parameter $k_BT_s|\cos\gamma|/v$ (see Fig. \ref{fig:thermal}). In the limit $T_s\cos\gamma \to 0$, the universal ratio \eqref{WF} is recovered, but, in the case $\gamma \neq \pi/2$, the Wiedemann-Franz law is in general violated for $T_s>0$ due to the boundary conditions and the behavior of the density of surface states. 

\begin{figure}[tb]
\includegraphics[width=\columnwidth]{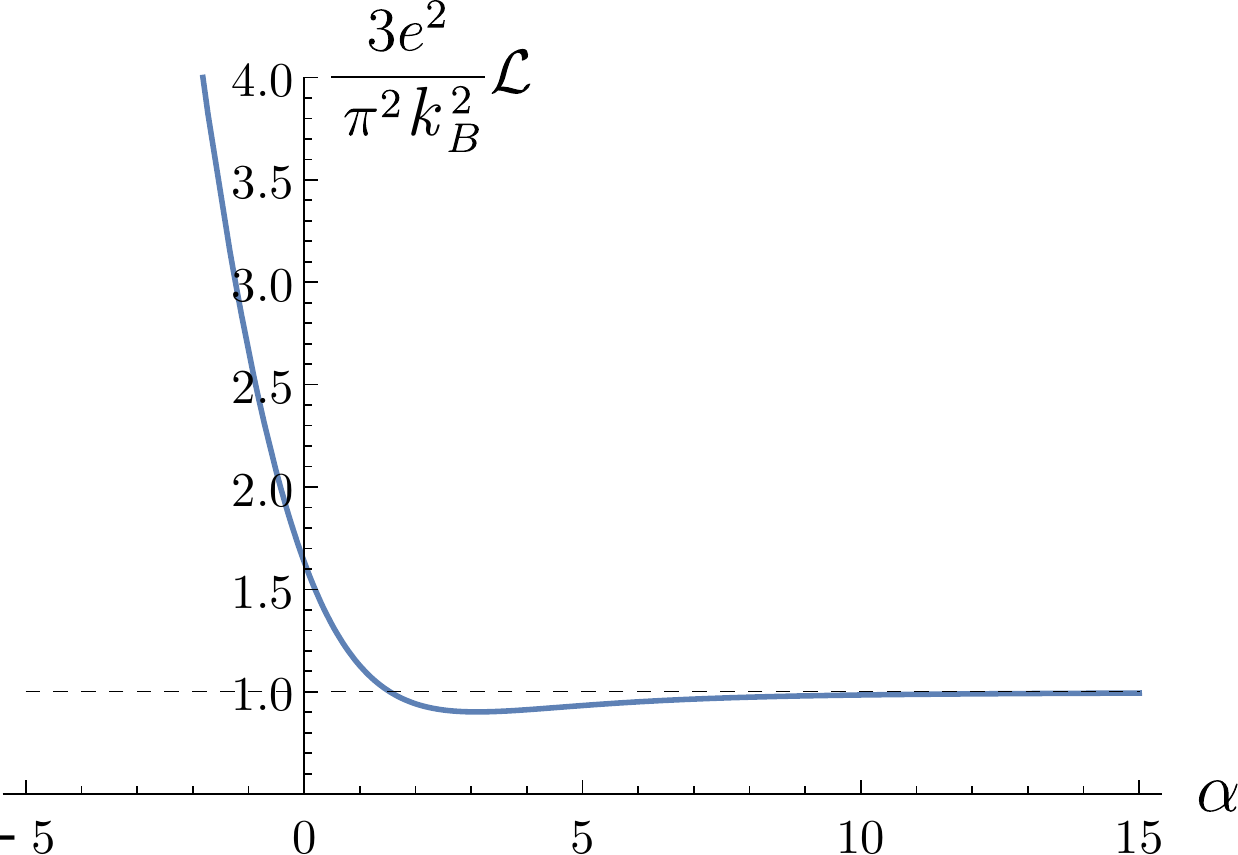}
\caption{Normalized Lorentz ratio $3e^2\mathcal{L}/\pi^2k_B^2$, derived from Eq. \eqref{Lorentzmaster}, as a function of the parameter $\alpha$ defined in Eq. \eqref{alpha}.}
\label{fig:lorentzuniv}
\end{figure}

\begin{figure}[tb]
\includegraphics[width=\columnwidth]{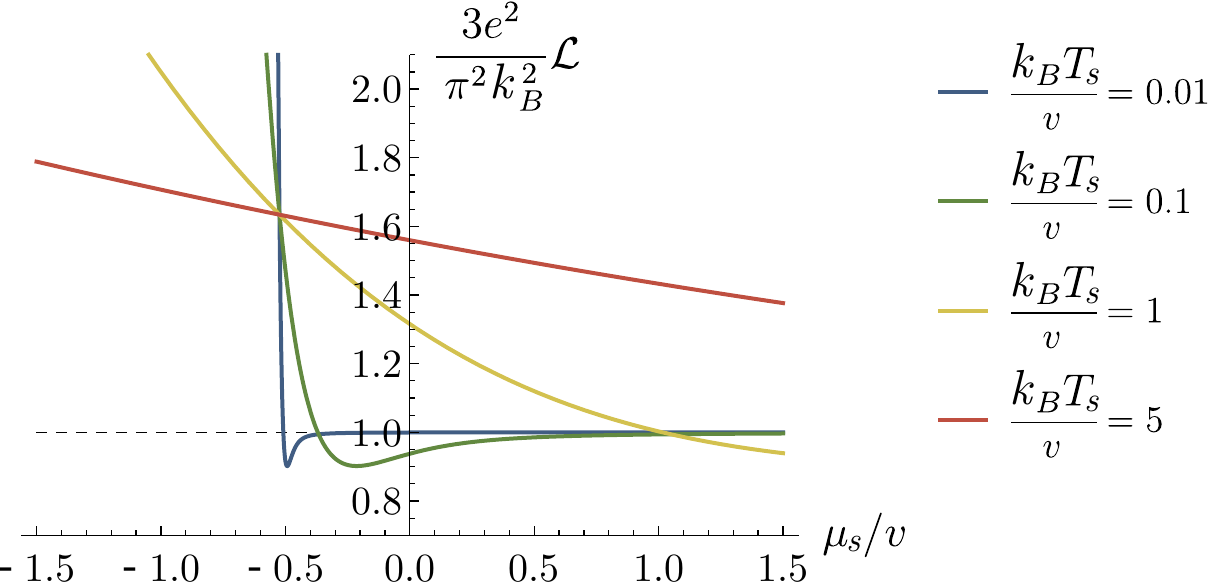}
\caption{Lorentz ratio $\mathcal{L}$ as a function of the chemical potential $\mu_s/v$. The Lorentz ratio is normalized by the parameter ${\pi^2k_B^2}/{3e^2}$ and is calculated from Eqs. \eqref{alpha} and \eqref{Lorentzmaster} for $\gamma=\pi/3$, $p_0=\pi/6$ and several values of the temperature. For $0<\gamma<\pi/2$, the Wiedemann-Franz law is violated in the low-temperature regime for chemical potentials sufficiently negative, as exemplified by the curve at $k_BT_s/v=0.01$.}
\label{fig:lorentzmu}
\end{figure}

For $\mu_s \neq 0$, $\kappa_H$ depends separately on $T_s$ and $\gamma$ and the Lorentz ratio displays a rich behavior (see Figures \ref{fig:lorentzgamma} and \ref{fig:lorentzmu}). From Eqs. \eqref{Hallgeneral} and \eqref{thermalhall} it is possible to derive that the Lorentz ratio depends solely on the parameter: 
\begin{equation}
\label{alpha}
\alpha \equiv v\beta_s\left[\frac{p_0}{2|\cos\gamma|}+\textrm{Sign}\left(\pi/2-\gamma\right)\frac{\mu_s}{v}\right]\,.
\end{equation}
For $\cos\gamma >0$, $\alpha=0$ when the chemical potential coincides with the minimum of the energy band of the surface states; for $\cos\gamma <0$ and $\alpha=0$, $\mu$ lies on the surface energy maximum instead. The values $\alpha>0$ thus correspond to the chemical potential lying within the energy band of the surface states; whereas for $\alpha<0$, the chemical potential lies outside. 

From the integral in Eq. \eqref{thermalhall} we derive:
\begin{equation}\label{Lorentzmaster}
\mathcal{L}(\alpha) = \frac{k_B^2}{4e^2}\left[4\alpha^2 -12\alpha\frac{{\rm Li}_{\frac{3}{2}}\left(-e^{\alpha}\right)}{{\rm Li}_{\frac{1}{2}}\left(-e^{\alpha}\right)} + 15\frac{{\rm Li}_{\frac{5}{2}}\left(-e^{\alpha}\right)}{{\rm Li}_{\frac{1}{2}}\left(-e^{\alpha}\right)} \right],
\end{equation}
where ${\rm Li}$ labels polylogarithm functions \cite{mathbooks}.

In Fig. \ref{fig:lorentzuniv} we illustrate the general behavior of $\mathcal{L}$ as a function of $\alpha$. The sign of the square bracket in \eqref{alpha} is particularly important because it determines the low-temperature behavior of the Lorentz ratio. In the limit $\alpha \to +\infty$, the Lorentz ratio converges to the standard value \eqref{WF} and the Wiedemann-Franz low is fulfilled for $T_s\to 0$; for $\alpha>0$, indeed, the curve in Fig. \ref{fig:lorentzuniv} reproduces the same results as Fig. \eqref{fig:thermal}. On the contrary, for negative values of $\alpha$, the Lorentz ratio increases and diverges as $k_B^2 \alpha^2/e^2$ for $\alpha\to -\infty$. For negative values of \eqref{alpha}, the surface states acquire indeed an insulating behavior in the low-temperature limit, since the chemical potential lies outside their energy band. The value at $\alpha=0$, instead, corresponds to the high-temperature limit for both the regimes. From Eq. \eqref{Lorentzmaster} we derive $\mathcal{L}(0)\approx 1.634$, which defines the high-temperature limit for all the values of $\gamma \neq \pi/2$ and all the values of $\mu_s$, compatibly with the regime of validity of the low-energy Hamiltonian \eqref{ham}. In this high-temperature limit, the Wiedemann-Franz law for the anomalous Hall conductivity is always violated for $\gamma \neq \pi/2$, despite the non-interacting nature of our model.

The dependence of the Lorentz ratio from the parameter $\alpha$ is reflected in the behaviors depicted in Figures \ref{fig:lorentzgamma} and \ref{fig:lorentzmu} when considering the thermal transport as a function of the boundary conditions and the chemical potential respectively. 

In Fig. \ref{fig:lorentzgamma} we consider the behavior of $\mathcal{L}$ for $\mu_s>0$ as a function of $\gamma$ for several values of the temperature: we immediately observe that there is a point, for $\gamma > \pi/2$, in which all the curves cross and the Lorentz ratio is independent of the temperature. This point coincides with $\alpha=0$ in Eq. \eqref{alpha}, thus for the chemical potential lying on the extremum of the surface energy band. For all the values of gamma on the left of this crossing point, we observe a non-monotonic behavior of $\mathcal{L}$ with the temperature, similar to the case of $\mu_s=0$ (Fig. \ref{fig:thermal}). These values of $\gamma$ span the range $\alpha>0$ in Fig. \ref{fig:lorentzuniv} and, in this regime, the Wiedemann-Franz law is recovered for $T \to 0$. The values of $\gamma$ larger than the crossing point, instead, span the regime $\alpha<0$ due to $\mu_s>0$. For decreasing temperature and for each value of $\gamma$ in this range, the Lorentz ratio diverges.

A corresponding behavior is obtained also as a function of $\mu_s$. Fig. \ref{fig:lorentzmu} illustrates the Lorentz ratio as a function of $\mu_s/v$ for different values of the temperature at $\gamma=\pi/3$. There is a crossing point in which $\mathcal{L}$ does not depend on the temperature. This is the value of $\mu_s$ such that $\alpha=0$ in Eq. \eqref{alpha}. On the left of this point the Lorentz ratio increases by lowering the temperature. On the right of the crossing point, instead, $\alpha>0$ and we recover the Wiedemann-Franz limit for $T_s \to 0$.

\section{Conclusions} \label{concl}

In this work we analyzed in detail the surface transport properties of a toy model of Weyl semimetal with two band-touching points, thus breaking time-reversal symmetry, which can be adopted, \textcolor{black}{for example,} to describe layered intermetallic materials with magnetically induced Weyl points \cite{soh2019}. \textcolor{black}{Additionally}, we expect that our results can be easily extended to account for multiple pairs of Weyl points, for materials in which such pairs are well-isolated in the Brillouin zone.

Our analysis stems from the self-adjoint extensions of the bulk Weyl Hamiltonian and allows us to focus on the non-universal properties of the transport that depend on the boundary conditions. In particular, the set of boundary conditions we considered is defined by a single angle $\gamma$ that determines both the pseudospin polarization at the surface of the system and the shape of the Fermi arcs.

From experimental data \cite{souma2016,morali2019} and \textit{ab initio} simulations \cite{sun2015} of Weyl semimetals, it is known that the lattice termination of the system (related to the pseudospin polarization) strongly affects the properties of the Fermi arcs. In full generality, the boundary spin polarization may depend on the conserved momenta of the system, thus it can rotate along the Fermi arcs in the surface Brillouin zone \cite{sun2015}. Our model considers instead a simpler case with a constant pseudospin polarization $\gamma$ independent of the momenta, as obtained by the self-adjoint Hamiltonian extensions with local boundary conditions. The polarization $\gamma$ we consider is not necessarily equivalent to the physical spin of the system, thus our model is compatible with former results. Despite its simplicity, our work demonstrates that it is possible to consistently include the boundary conditions in a field theoretical determination of the surface transport properties.

We studied in detail the Fermi arcs of the system and their contribution to the anomalous Hall conductivity by calculating the corresponding current density, which typically decays with the square of the distance from the surface.

We derived a general formula for the anomalous Hall conductivity as a function of temperature, chemical potential and boundary parameter for both the electric and thermal transport. This allowed us to verify that, in general, the Wiedemann-Franz law for the anomalous Hall transport is fulfilled only in the very limit $T \to 0$, whereas the Lorentz ratio presents a non-trivial behavior as a function of chemical potential and boundary conditions for finite temperatures.

We additionally estimated the bulk conductance of our two-Weyl-point model based on a Landauer-B\"uttiker approach, and we verified with numerical calculations our predictions at zero temperature based on a suitable lattice model. In particular, we show that, also in lattice systems, it is possible to vary the boundary conditions through the introduction of suitable surface Zeeman interactions. Also in this case, the boundary polarization, the shape of the Fermi arcs and the density of surface states are linked to the same parameter $\gamma$.

\section*{Acknowledgements} We thank Ajit C. Balram, Ion Cosma Fulga, Pietro Novelli, Marco Polini, and Fabio Taddei for useful discussions and the developers of {\it Kwant}. M. B. was supported by a research grant (Project nr. 25310) from Villum Fonden.

\appendix   

\section{The eigenvectors of the self-adjoint extensions $H_\gamma$} 

We describe here the explicit form of the $H_\gamma$-eigenvectors and 
summarize their basic properties. The surface eigenstates are given by 
\bwt
\be
\zeta_{\rm s}^\pm ({\bf r}, p) = 
\Theta[\pm \varepsilon_{\rm s}(p)]\Theta[\tilde{p}_z(p)] \sqrt {2\tilde{p}_z(p)}\e^{-z \tilde{p}_z(p)+\ri(xp_x+yp_y)} w(\gamma)\, , \qquad 
w(\gamma)=\frac{1}{\sqrt 2}\begin{pmatrix} \e^{-\ri \gamma/2}\\ \e^{\ri \gamma/2}\end{pmatrix}\, , 
\label{A1}
\ee
\ewt 
where $\varepsilon_{\rm s}(p)$ and $\tilde{p}_z(p)$ are defined by (\ref{s2}) and (\ref{g}). The exponential decay along the $z$-axis is worth mentioning. 

For the bulk eigenstates one has 
\bwt
\be
\zeta_{\rm b}^\pm ({\bf r, p}) = 
\Theta[p_z] \left [\e^{\ri {\bf p}_{\rm r}{\bf r}} u^{\pm}({\bf p}_{\rm r})+ S_\pm ({\bf p})\e^{\ri {\bf p} {\bf r}} u^{\pm}({\bf p})\right ]\, , 
\qquad {\bf p}_{\rm r} = (p_x,p_y,-p_z)\, ,  
\label{A2}
\ee
\be
u^+({\bf p})=n({\bf p})\begin{pmatrix} vp_z+\varepsilon_{\rm b}({\bf p})\\ v [\ri p_y+g(p_x)]\end{pmatrix}\, , \quad 
u^-({\bf p})=n({\bf p})\begin{pmatrix} v [\ri p_y-g(p_x)]\\ vp_z-\varepsilon_{\rm b}({\bf p})\end{pmatrix}\, , 
\qquad n({\bf p}) = \frac{v}{\sqrt{\varepsilon_{\rm b}({\bf p})[\varepsilon_{\rm b}({\bf p})+v p_z]}}\, , 
\label{A3}
\ee
where $\varepsilon_{\rm b}({\bf p})$ is defined by (\ref{b2}) and 
\be
S_+ (\bm p) = \left [ \frac{\varepsilon_{\rm b}({\bf p}) +v p_z}{\varepsilon_{\rm b}({\bf p}) - v p_z} \right ]^{1/2}  \frac{vp_z - 
\varepsilon_{\rm b}({\bf p}) + v[g(p_x) + i p_y] e^{-\ri  \gamma}}{vp_z + 
\varepsilon_{\rm b}({\bf p}) - v[g(p_x) + i p_y] e^{-\ri  \gamma}} \; , 
\label{A4}
\ee
\be
S_- (\bm p) = \left [ \frac{\varepsilon_{\rm b}({\bf p}) +v p_z}{\varepsilon_{\rm b}({\bf p}) - v p_z} \right ]^{1/2}  \frac{vp_z - 
\varepsilon_{\rm b}({\bf p}) - v[g(p_x) - i p_y]e^{\ri  \gamma} }{vp_z + 
\varepsilon_{\rm b}({\bf p}) + v[g(p_x) - i p_y]e^{\ri  \gamma} } \; . 
\label{A5}
\ee
\ewt 
The factors (\ref{A4},\ref{A5}) satisfy 
\be
|\, S_{\pm } (\bf p ) \, | = 1 
\label{A6}
\ee 
and have a simple physical interpretation: they represent the scattering matrices for particles and antiparticles 
with incoming momenta ${\bf p}_{\rm r}$, which 
are reflected from the boundary $z =0$ and have final momenta $\bf p$. 

The fundamental property of the surface and bulk states is that they form a complete system  
\bwt
\be
\sum_{\sigma = \pm }\left [\int \frac{d^3p}{(2 \pi)^3} \, \zeta_{\rm b}^\sigma({\bf r, p})^\dagger_\beta \, \zeta_{\rm b}^\sigma({\bf r^\prime, p})_\alpha + 
 \int \frac{d^2p}{(2 \pi)^2}  \, \zeta_{\rm s}^\sigma({\bf r}, p)^\dagger_\beta \, \zeta_{\rm s}^\sigma({\bf r}^\prime, p)_\alpha \right ] =  
 \delta_{\alpha \beta} \, \delta^3 ({\bf r - r^\prime} ) \; , \qquad z,z^\prime >0\, , 
\label{A7}
\ee
\ewt 
which is the main ingredient for constructing the canonical quantum field $\Psi$ defined by (\ref{tfield},\ref{s3},\ref{b3}). 
For proving (\ref{A7}) one can proceed as follows. One starts by considering the three-dimensional integral in the square brackets, 
which gives the right hand side plus a rest. The latter can be reduced 
to a two-dimensional integral by integrating over $p_z$, using the Cauchy integral formula, $z>0$ and the analytic properties of the scattering 
matrices $S_{\pm } (\bf p )$ in the complex $p_z$-plane. At this point the rest precisely cancels the second term in the square brackets. 
This computation is very instructive because it shows that the surface states are generated by the bound states (poles in the upper 
half complex $p_z$-plane) of the scattering matrices (\ref{A4},\ref{A5}). \\

\section{Estimation of the bulk conductance} \label{app:conductance}

For the evaluation of Eq. \eqref{transmission} we consider wavefunctions at $E=0$ of the kind:
\bwt
\begin{align}
&\psi_S(y<0) = \frac{1}{\sqrt{2}}\begin{pmatrix} i \\ 1 \end{pmatrix} \e^{i \left(p_x x +p_y y + p_z z\right)} + \frac{r}{\sqrt{2}}\begin{pmatrix} -i \\ 1 \end{pmatrix} \e^{i \left(p_x x -p_y y + p_z z\right)} \\
&\psi(0<y<L) = \alpha_+ \begin{pmatrix} i \tilde{p}_y -g \\ p_z \end{pmatrix} \e^{i \left(p_x x +\tilde{p}_y y + p_z z\right)} + \alpha_- \begin{pmatrix} -i \tilde{p}_y -g \\ p_z \end{pmatrix} \e^{i \left(p_x x -\tilde{p}_y y + p_z z\right)}\\
&\psi_D(y>L) = \frac{t}{\sqrt{2}}\begin{pmatrix} i \\ 1 \end{pmatrix} \e^{i \left(p_x x +p_y (y-L) + p_z z\right)}\,.
\end{align}
\ewt
Here, $S$ and $D$ label the source and drain external leads; $r$ and $t$ are the reflection and transmission amplitudes, such that $|t|^2+|r|^2=1$, and they depend on $p_z$ and $g(p_x)$. The spinors in the leads are polarized along the $\hat{y}$ axis, consistently with the approximation $\mu_{\rm lead} \to \infty$ and $|p_y| \gg |p_x|,|p_z|$. The chemical potential in the central region is $\mu=0$ and
\begin{equation}
\tilde{p}_y= i\sqrt{h^2+p_z^2}\,,
\end{equation}
due to the vanishing energy.  To estimate the transmission probability $\mathcal{T}=|t|^2$ we impose the specific boundary conditions:
\begin{align}
&\psi_S(x,y=0^-,z) = \psi(x,y=0^+,z) \,,\\
&\psi(x,y=L^-,z) = \psi_D(x,y=L^+,z)\,;
\end{align}
we obtain:
\begin{equation}
|r|^2 = \tanh^2\left(L \sqrt{g(p_x)^2+p_z^2}\right)\,,
\end{equation}
from which we derive Eq. \eqref{transmission}. A more rigorous estimate of the transmission coefficient should take into account more general boundary conditions at $y=0$ and $y=L$, which, also in this case, could be classified through the self-adjoint extensions of the full Hamiltonian of the system.

\end{document}